\documentclass[useAMS,usenatbib]{mn2e}

\usepackage{graphicx}
\usepackage{epsfig}
\usepackage{natbib}
\usepackage{times}
\usepackage{ulem}
\usepackage[T1]{fontenc}
\newcommand{\fesc}{\ifmmode{f_{\rm esc}}\else{$f_{\rm esc}$}\fi}
\newcommand{\fescs}{\ifmmode{f_{\rm esc}^\star}\else{$f_{\rm esc}^\star$}\fi}
\newcommand{\kms}{\ifmmode{{\;\rm km~s^{-1}}}\else{km~s$^{-1}$}\fi}
\newcommand{\fgas}{\ifmmode{{f_{\rm gas}}}\else{$f_{\rm gas}$}\fi}
\newcommand{\cubecm}{\ifmmode{{\rm cm^{-3}}}\else{cm$^{-3}$}\fi}
\newcommand{\ztwo}{\ifmmode{{\rm [Z_2/H]}}\else{[Z$_2$/H]}\fi}
\newcommand{\zthree}{\ifmmode{{\rm [Z_3/H]}}\else{[Z$_3$/H]}\fi}
\newcommand{\lsim}{\lower0.3em\hbox{$\,\buildrel <\over\sim\,$}}
\newcommand{\gsim}{\lower0.3em\hbox{$\,\buildrel >\over\sim\,$}}

\newcommand{\sfr}{\ifmmode{\textrm{M}_\odot \,\textrm{yr}^{-1} \,\textrm{Mpc}^{-3}}\else{M$_\odot$ yr$^{-1}$ Mpc$^{-3}$}\fi}
\newcommand{\hsfr}{\ifmmode{\textrm{M}_\odot\, \textrm{yr}^{-1}}\else{M$_\odot$ yr$^{-1}$}\fi}

\newcommand{\eavg}{\ifmmode{\langle E_\gamma \rangle}\else{$\langle E_\gamma \rangle$}\fi}

\newcommand{\enzo}{{\sc enzo}}
\newcommand{\yt}{{\sc yt}}
\newcommand{\moray}{{\sc enzo+moray}}
\newcommand{\Ms}{\ifmmode{M_\odot}\else{$M_\odot$}\fi}
\newcommand{\vrms}{\ifmmode{v_{\rm rms}}\else{$v_{\rm rms}$}\fi}

\newcommand{\hh}{H$_2$}

\newcommand{\tvir}{\ifmmode{T_{\rm{vir}}}\else{$T_{\rm{vir}}$}\fi}
\newcommand{\mvir}{\ifmmode{M_{\rm{vir}}}\else{$M_{\rm{vir}}$}\fi}
\newcommand{\rvir}{\ifmmode{r_{\rm{vir}}}\else{$r_{\rm{vir}}$}\fi}

\newcommand{\lya}{Ly$\alpha$}
\newcommand{\jj}{\ifmmode{J_{21}}\else{$J_{21}$}\fi}
\newcommand{\flw}{\ifmmode{F_{LW}}\else{$F_{LW}$}\fi}
\newcommand{\kph}{\ifmmode{k_{\rm ph}}\else{$k_{\rm ph}$}\fi}

\newcommand{\zsun}{\ifmmode{\rm\,Z_\odot}\else{$\rm\,Z_\odot$}\fi}

\newcommand{\hi}{H {\sc i}}
\newcommand{\hii}{H {\sc ii}}
\newcommand{\hei}{He {\sc i}}
\newcommand{\heii}{He {\sc ii}}
\newcommand{\heiii}{He {\sc iii}}
\newcommand{\nhi}{\ifmmode{N_{\rm HI}}\else{$N_{\rm HI}$}\fi}
\newcommand\unit[1]{\; \textrm{#1}}
\newcommand{\pem}{\unit{s}^{-1} \unit{cMpc}^{-3}}

\def\eps@scaling{1.0}%
\newcommand\epsscale[1]{\gdef\eps@scaling{#1}}%
\newcommand\plotone[1]{%
 \centering 
 \leavevmode 
 \includegraphics[width={\eps@scaling\columnwidth}]{#1}%
}%
\newcommand\plottwo[2]{%
 \centering 
 \includegraphics[width={\eps@scaling\columnwidth}]{#1}%
 \hfil 
 \includegraphics[width={\eps@scaling\columnwidth}]{#2}%
}%

\begin{document}

\title[Propelling reionisation with the faintest galaxies]{The birth
  of a galaxy -- III. Propelling reionisation with the faintest
  galaxies}

\author[J. H. Wise et al.]{John H. Wise$^1$\thanks{e-mail:
    jwise@physics.gatech.edu}, Vasiliy G. Demchenko$^1$, Martin T.
  Halicek$^1$, Michael L. Norman$^2$, \newauthor Matthew J. Turk$^3$,
  Tom Abel$^4$, and Britton D. Smith$^5$
  \vspace{0.4cm}\\
  $^{1}$ Center for Relativistic Astrophysics, Georgia Institute of
  Technology, 837 State Street, Atlanta, GA
  30332, USA\\
  $^{2}$ Center for Astrophysics and Space Sciences, University
  of California at San Diego, La Jolla, CA 92093, USA\\
  $^{3}$ Department of Astronomy, Columbia University, 538 West 120th
  Street, New York, NY 10027, USA\\
  $^{4}$ Kavli Institute for Particle Astrophysics and Cosmology,
  Stanford University, Menlo Park, CA 94025, USA\\
  $^{5}$ Institute of Astronomy, University of Edinburgh, Blackford
  Hill, Edinburgh EH9 3HJ, UK}

\pagerange{\pageref{firstpage}--\pageref{lastpage}} \pubyear{2014}

\maketitle
\label{firstpage}

\begin{abstract}

  Starlight from galaxies plays a pivotal role throughout the process
  of cosmic reionisation.  We present the statistics of dwarf galaxy
  properties at $z > 7$ in haloes with masses up to $10^9 \Ms$, using
  a cosmological radiation hydrodynamics simulation that follows their
  buildup starting with their Population III progenitors.  We find
  that metal-enriched star formation is not restricted to atomic
  cooling ($\tvir \ge 10^4 \unit{K}$) haloes, but can occur in haloes
  down to masses $\sim10^6 \Ms$, especially in neutral regions.  Even
  though these smallest galaxies only host up to $10^4 \Ms$ of stars,
  they provide nearly 30 per cent of the ionising photon budget.  We
  find that the galaxy luminosity function flattens above $M_{\rm UV}
  \sim -12$ with a number density that is unchanged at $z \la 10$.
  The fraction of ionising radiation escaping into the intergalactic
  medium is inversely dependent on halo mass, decreasing from 50 to 5
  per cent in the mass range $\log M/\Ms = 7.0-8.5$.  Using our galaxy
  statistics in a semi-analytic reionisation model, we find a Thomson
  scattering optical depth consistent with the latest {\it Planck}
  results, while still being consistent with the UV emissivity
  constraints provided by \lya~forest observations at $z = 4-6$.

\end{abstract}

\begin{keywords}
  cosmology: reionisation -- galaxies: formation -- galaxies: dwarf --
  galaxies: high-redshift -- methods: numerical -- radiative transfer
\end{keywords}

\section{Introduction}








Cosmic reionisation is an extended process as individual \hii~regions
grow around ionising sources that gradually coalesce, culuminating in
a fully ionised universe by $z \sim 6$ \citep[e.g.][]{Gnedin97,
  Razoumov02, Sokasian03, Ciardi03, Furlanetto04, Iliev06,
  Robertson10, Zahn11, Trac11, So13}.  However, there is still some
tension between observational constraints on the timing and duration
of reionisation.  First, the transmission fraction of $z \sim
  6$ quasar light blueward of \lya~through the intergalactic medium
  (IGM) indicates that the universe was mostly ionised by this epoch
\citep[e.g.][]{Gunn65, Fan02, Fan06_QSO, Willott07, Mortlock11}.
Second, observations of the cosmic microwave background (CMB) from the
{\it Wilkinson Microwave Anisotropy Probe (WMAP)} and {\it Planck}
have measured the optical depth to Thomson scattering $\tau_e =
0.089^{+0.012}_{-0.014}$, which corresponds to the universe being
$\sim$50 per cent ionised at $z = 11.1 \pm 1.1$
\citep{Planck13_Cosmo}.  But the ionising emissivity measured at $z =
4-6$ through \lya~forest observations cannot account for this measured
$\tau_e$, indicating that the end of reionisation must be
photon-starved \citep{Bolton07} and that the emissivity must have been
higher during reionisation.  Third, the duration%
\footnote{\citet{Zahn12} defines $\Delta z$ as the redshift elapsed
  between 20 and 99 per cent ionised.} %
of reionisation has been constrained to occur within $\Delta z < 7.9$
by measuring the kinetic Sunyaev-Zel'dovich effect with the {\it South
  Pole Telescope} \citep[SPT;][]{Zahn12}.  These observations suggest
that reionisation was an extended process, mainly occurring at $6 \la
z \la 15$.

What population of ionising sources drives this global and extended
transition?  It is clear that quasars and the very brightest galaxies,
both of which are too rare, do not significantly contribute to the
overall ionising photon budget of reionisation
\citep[e.g.][]{Shapiro86, Dijkstra04, Willott10, Grissom14}.
Starlight from galaxies are thought to provide the vast majority of
the ionising photon budget from extrapolating the observed $z > 6$
galaxy luminosity function (LF) to low luminosities
\citep[e.g.][]{Madau99, Bouwens12_Reion, Haardt12, Shull12,
  Fontanot12, Robertson13}.  Alternatively, massive, metal-free
(Population III; Pop III) stars can contribute on the order of 10 per
cent of the budget \citep{Ricotti04_P3, Greif06, Trac07, Wise08_Reion,
  Ahn12, Wise12_Galaxy, Paardekooper13, Johnson13} because they are
short-lived \citep[e.g.][]{Tumlinson00, Schaerer02} and can be
suppressed by chemical enrichment and \hh-dissociating radiative
feedback \citep[e.g.][]{Haiman97, Haiman00, Machacek01, Wise08_Gal}.
Finally, X-ray radiation from X-ray binaries and accreting massive
black holes partially ionise the IGM and may contribute a small amount
to the Thomson scattering optical depth \citep{Ricotti04_Xray,
  McQuinn12, Power13, Fragos13}.

Deep galaxy surveys, such as the {\it Hubble Ultra Deep Field}
\citep[HUDF;][]{Beckwith06, Koekemoer13} and CANDELS \citep{CANDELS,
  Koekemoer11}, can probe $z \ga 6$ galaxies up to absolute UV
magnitudes $M_{\rm UV} < -18$ or equivalently a stellar mass $M_\star
\ga 10^8 \Ms$.  Future deep surveys using the {\it James Webb Space
  Telescope (JWST)} and 30-m class ground-based telescopes will push
this limit down to $M_{\rm UV} \sim -15.5$.  At these high redshifts,
the faint-end LF slope is steepening with redshift and is around $-2$
at $z \sim 8$ \citep[e.g.][]{Bouwens11, Bradley12, Oesch12_LF}.  The
least massive galaxies can be suppressed through supernovae (SNe)
and/or radiative feedback, and this process should materialise as a
flattening or turn-over in the LF, but starting at which limiting
magnitude?

Any change in the LF behavior should be related to the star formation
efficiency, i.e. $M_\star/M_{\rm gas}$, which is also connected to the
effectiveness of gas cooling and the halo mass in principle.  For
instance, starting at the lowest mass haloes, the primary coolants in
the interstellar medium (ISM) are molecular hydrogen and metals
(e.g. C, O, Si), whereas above a virial temperature $\tvir \sim 10^4
\unit{K}$, they change to atomic hydrogen.  Negative feedback in these
small haloes is also a concern on whether they can sustain efficient
star formation.  Examples of such feedback include photo-evaporation,
\hh~dissociation, and gas blowout from \hii~regions and SN, which all
depend on halo mass \citep[e.g.][]{Gnedin00, Wise09, Stinson13,
  Hopkins13_FIRE}.  In the \hh-cooling haloes with $M \sim 10^6 \Ms$,
\hh~can be dissociated by a moderate Lyman-Werner (LW; 11.2--13.6~eV)
background.  But in more massive haloes with $M \sim 10^7 \Ms$, cold
gaseous reservoirs can form even in the presence of a strong LW
radiation field \citep{Wise07_UVB, Johnson08, OShea08,
  Safranek12}.  Progressing up the mass scale, even atomic cooling
haloes are prone to negative feedback; for example, haloes with masses
$M \lsim 2 \times 10^9 \Ms$ can be photo-evaporated by an external
radiation field, gradually boiling away their star-forming gas
reservoir \citep{Efstathiou92, Thoul96, Dijkstra04, Shapiro04,
  Okamoto08, Finlator11}.

How does the LF limiting magnitude $M_{\rm lim}$ affect reionisation
histories?  Recently, a few groups \citep{Kuhlen12, Finkelstein12,
  Robertson13} have explored this question among other variations in
their reionisation models.  \citeauthor{Kuhlen12} found a nearly
linear dependence between $\tau_e$ and $M_{\rm lim}$, increasing
$\tau_e$ from $\sim$0.06 to 0.08 when $M_{\rm lim}$ increases from
--13 to --10.  \citeauthor{Robertson13} found little dependence on
$M_{\rm lim}$ above --13.  Whereas \citeauthor{Finkelstein12} showed
that the escape fraction must be greater than 30 and 50 per cent to
sustain reionisation at redshifts 6 and 7, respectively, if only the
observed CANDELS galaxies contribute to the emissivity.  Additional
constaints on reionisation can be gained from the inferred ionising
flux in the \lya~forest at $z = 4-6$ \citep{Bolton07, Kuhlen12}.
Extrapolating the LF down to $M_{\rm UV} = -13$ and assuming that the
escape fraction \fesc~is independent of halo mass,
\citeauthor{Finkelstein12} also found that $\fesc < 0.13$ is
constrained by the measured ionising photon emissivity in \lya~forest
observations at $z = 6$.  It is clear from these studies that a
population of unobserved dwarf galaxies are primarily responsible for
driving cosmic reionisation.

To further refine reionisation models, it is pertinent to determine
the characteristic properties of these unobserved dwarf galaxies.  In
particular, the stellar fraction, $M_\star/M_{\rm vir}$, and \fesc~of
high-redshift galaxies are the largest sources of uncertainty in
reionisation models\footnote{Other properties that could affect the
  ionising emissivity originating from such galaxies are the gas
  fraction, initial mass function (IMF), gaseous and stellar
  morphology, and the strength and duration of star formation
  events.}.  Current reionisation models favor scenarios that have a
luminosity-weighted average escape fraction that increases with
redshift to match the observed $\tau_e$ value while being
photon-starved at $z \sim 6$ \citep{Alvarez12, Haardt12, Kuhlen12,
  Shull12, Mitra13}.  For example, the model of \citeauthor{Alvarez12}
considers a scenario where galaxies in haloes of mass $10^8 \le M/\Ms
\le 2 \times 10^9$ with $\fesc = 0.8$ dominate the ionising emissivity
at early times and are gradually photo-suppressed \citep[also
see][]{Sobacchi13}.  Then at $z \la 6.5$, galaxies greater than $2
\times 10^9 \Ms$ with lower average escape fractions become
sufficiently abundant to produce the majority of ionising photons,
keeping the universe ionised in a photon-starved scenario.  Before
moving forward, it should be stressed that the UV escape fraction is
an intrinsic quantity for a given galaxy not an entire population.
Galaxies with the same mass can have very different escape fractions,
arising from, e.g., complex gaseous and stellar morphologies, dust
content, and cosmological mass inflow.  Furthermore, variable star
formation rates (SFRs) and the associated radiative feedback in the
ISM can result in an escape fraction that is highly time-dependent.

The ionising escape fraction is a notoriously difficult quantity to
measure both in high-redshift galaxy observations and theoretical
studies.  Nevertheless, this topic has been a subject of great
interest to constrain the reionisation history of the universe.  On
the observational side, it is nearly impossible to detect Lyman
continuum (LyC) emission at $z > 4$ because the number density of
Lyman limit systems rapidly increases with redshift \citep{Inoue08}.
However at $z \sim 3$ when the IGM optical depth is around unity,
detection of intrinsic LyC radiation becomes feasible.  Deep
narrow-band galaxy imaging and spectroscopy have uncovered LyC
emission in 10--20 per cent of Lyman-break galaxies, which can be
interpreted as the mean ionising radiation escape fraction
\citep{Steidel01, Shapley06, Iwata09, Nestor11, Jones13}, but see
\citet{Vanzella12}.

Theoretical efforts have focused on the calculation of \fesc~for over
a decade, using analytical and numerical techniques with varying model
complexities.  The galaxies studied in these cumulative works span
over six orders of magnitude in halo mass and are considered out to $z
= 15$.  Models of the escape fraction found that $\fesc \la 0.06$ for
Milky Way like galaxies \citep{Dove00}, and, in general, it depends on
the density structure of the ISM and SFR \citep{Ciardi02, Clarke02,
  Wise09, Fernandez11, Benson13}.  At high redshift, \citet{Ricotti00}
found higher escape fractions $\fesc \ga 0.1$ in haloes with masses $M
\le 10^7 \Ms$, but they posed the valid question of whether these
low-mass haloes can host star formation.  This paper will address this
exact question, utilising a cosmological radiation hydrodynamics
simulation of dwarf galaxy formation.

Conversely, because of the higher mean densities at high redshift,
\citet{Wood00} argued that $\fesc \le 0.01$, and \citet{Fujita03}
found that $\fesc \le 0.1$ from dwarf starburst disc galaxies with
total masses between $10^8$ and $10^{10} \Ms$.  \citet{Paardekooper11}
found similar results for isolated high-redshift disc galaxies with
total masses of $10^8$ and $10^9 \Ms$.  All of the aforementioned
models were idealised calculations of isolated galaxies; however
\citet{Razoumov06, Razoumov07} and \citet{Gnedin08} used cosmological
simulations of galaxy formation with radiative transfer to conclude
that $\fesc = 0.01-0.1$ and $\fesc \sim 0.01 - 0.03$, respectively, in
haloes with $M \ge 10^{11} \Ms$ at $z = 3-5$.  If these low escape
fractions were present in lower mass galaxies before reionisation,
insufficient LyC emission would escape from them to reionise the
universe by $z = 6$ \citep{Gnedin08L}.

In radiation hydrodynamics simulations of isolated dwarf irregular
galaxies at $z = 8$, \citet{Wise09} found that $\fesc \ga 0.3$, which
was confirmed by several other groups with numerical simulations
shortly afterward \citep{Razoumov10, Yajima11, Paardekooper13,
  Ferrara13}.  These works imparted momentum to the idea that
protogalaxies could be the dominant driver of reionisation, as
originally proposed by \citet{Ricotti00}.  Semi-analytic models of
reionisation tested these ideas and further constrained the required
escape fraction in high-redshift dwarf galaxies to be increasing with
redshift, suggesting that low-mass galaxies with high
\fesc~contributed a significant amount of the ionising photon budget
\citep{Haardt12, Alvarez12, Mitra13}.


Unfortunately, not even {\it JWST} has the capability to directly
detect the lowest-luminosity galaxies that could provide the majority
of ionising photons during the earlier epochs of reionisation.
Comparisons to local dwarf galaxies can be made, but in principle,
similarities could be few because some form in a neutral and cool
environment, largely unaffected by the ensuing inhomogeneous
reionisation.  In addition, some are directly affected by radiative
and SN feedback from Pop III stars \citep{Johnson06, Wise08_Gal,
  Greif10, Maio10_Pop32, Wise12_Galaxy, Pawlik13, Muratov13}.  Thus
for the time being, this problem is best approached theoretically.


This paper focuses on these early dwarf galaxies that are sensitive to
feedback effects in haloes with $M \lsim 10^9 \Ms$.  The primary goal
of this paper is to quantify the mean stellar and gaseous properties,
the ionising escape fractions, and LFs of high-redshift dwarf
galaxies, and their contribution to global reionisation.  In the next
section, we outline our simulation set-up and methods.  Then, in
Section 3, we present scaling relations for stellar mass, gaseous
fractions, intrinsic UV magnitudes, and ionising escape fractions of
the simulated dwarf galaxies.  Next, in Section 4, we apply our mean
scaling relations to a semi-analytic reionisation model and show the
resulting reionisation history when low-luminosity galaxies are
considered.  We discuss the implications of our results and possible
observational signatures of the first galaxies in Section 5, where we
also compare our results to previous studies.  Lastly, we summarise
our findings in Section 6.

\section{Methods}
\label{sec:setup}

We further analyze the ``RP'' simulation originally presented in
\citet[][hereafter W12]{Wise12_RP}, which focused on the role of
radiation pressure during dwarf galaxy formation.  In this paper, we
will focus on the LF, escape fraction of UV radiation, and the role of
these first galaxies during reionisation.  A detailed description of
the radiative cooling, star formation, and stellar feedback models is
given in W12, thus we only briefly describe the input physics and
methods in this section.

\subsection{Simulation setup}
\label{sec:sim}

This simulation was run with the adaptive mesh refinement (AMR) code
\enzo~v2.0\footnote{\texttt{enzo-project.org, changeset 03a72f4f189a}}
\citep{Enzo}.  It uses an $N$-body adaptive particle-mesh solver
\citep{Efstathiou85, Couchman91, BryanNorman1997} to follow the DM
dynamics.  It solves the hydrodynamics equations using the
second-order accurate piecewise parabolic method \citep{Woodward84,
  Bryan95}, while an HLLC Riemann solver ensures accurate shock
capturing with minimal viscosity.  We use the nine-species (\hi, \hii,
\hei, \heii, \heiii, e$^-$, \hh, \hh$^+$, H$^-$) non-equilibrium
chemistry model in \enzo~\citep{Abel97, Anninos97} and the \hh~cooling
rates from \citet{Glover08_Rates}.  In addition to the primordial
radiative cooling, we include metal and molecular line cooling, using
cooling rates that are calculated with \textsc{cloudy}
\citep{Smith08_Cooling}.

We initialise a simulation of 1 comoving Mpc on a side at $z = 130$
with a base resolution of $256^3$, using \textsc{grafic}
\citep{Bertschinger01} with the 7-year \textsl{WMAP}
$\Lambda$CDM+SZ+LENS best fit \citep{WMAP7}: $\Omega_M = 0.266$,
$\Omega_\Lambda = 0.734$, $\Omega_b = 0.0449$, $h = 0.71$, $\sigma_8 =
0.81$, and $n = 0.963$ with the variables having their usual
definitions.  The DM mass resolution is 1840 \Ms, and the maximum
spatial resolution is 1.0 comoving pc or 12 levels of AMR refinement.
We refine the grid on baryon overdensities of $3 \times 2^{-0.2l}$,
where $l$ is the AMR level.  We also refine on a DM overdensity of
three and always resolve the local Jeans length by at least four
cells, avoiding artificial fragmentation during gaseous collapses
\citep{Truelove97}.  If any of these criteria are met in a single
cell, it is flagged for further spatial refinement.  We stop the
simulation at $z = 7.3$ when the simulation box is 70 per cent ionised
by volume, and the ray tracing from $\sim 1000$ point sources becomes
computationally expensive in the optically thin regime.  We analyze
the simulation from outputs that are written 12.5~Myr apart until $z =
8$ and then every 1~Myr until the final redshift.

\subsection{Star formation and feedback}
\label{sec:sf}

We use distinct star formation and feedback models for Population II
and III with the latter forming when the gas metallicity of the star
forming gas is below $10^{-4} \zsun$.  The star formation criteria are
similar to the original \citet{Cen92} model but with a critical
\hh~fraction for Population III star formation, and for Population II
stars, the model has been modified to allow for star-forming clouds
that are Jeans resolved.  We model the formation, main sequence, and
stellar endpoints of these populations with the following
characteristics:
\begin{itemize}
\item \textbf{Population II:} Each star particle represents a stellar
  cluster with a minimum mass M$_{\rm min} = 1000~\Ms$ and a Salpeter
  IMF.  The initial accretion event converts 7 per cent of the cold
  ($T < 1000$ K) gas within a sphere with a dynamical time of 3 Myr
  ($\bar{\rho} \simeq 1000\mu\, \cubecm$) into a star particle (see
  W12).  If the initial star particle is less than 1000~\Ms, then the
  star particle continues to accrete until it reaches this threshold.
  In the case where it does not reach $M_{\rm min}$ within a dynamical
  time of 3 Myr, the accretion halts at this point, and the star
  particle begins to radiate.  Star particles only emit radiation
  after this accretion phase is completed. They emit 6000 hydrogen
  ionising photons per stellar baryon over a lifetime of 20 Myr.  We
  consider a monochromatic spectrum with an energy of 21.6 eV
  \citep{Schaerer03}.  After living for 4 Myr, they generate $6.8
  \times 10^{48}$ erg s$^{-1}$ $\Ms^{-1}$ in SNe energy that are
  injected into resolved spheres of radius 10 pc.
\item \textbf{Population III:} Each star particle represents a single
  star whose mass is randomly drawn from a power-law IMF that
  exponentially decays below a characteristic mass $M_{\rm char} = 100
  \Ms$,
  \begin{equation}
    \label{eqn:imf}
    f(\log M) dM = M^{-1.3} \exp\left[-\left(\frac{M_{\rm
            char}}{M}\right)^{1.6}\right] dM.
  \end{equation}
  The mass-dependent hydrogen ionising and Lyman-Werner photon
  luminosities and lifetimes are taken from \citet{Schaerer02}.  We
  consider a monochromatic spectrum with an energy of 29.6 eV,
  appropriate for the near-constant $10^5$~K surface temperatures of
  Pop III stars.  The SN explosion energies and metal ejecta
  masses from \citet{Nomoto06} and \citet{Heger03} are used for Type
  II (20--40 \Ms) and pair instability SNe (140--260 \Ms),
  respectively.  We do not consider any SN feedback from Pop
  III stars in the mass range 40--140 \Ms~because they most likely
  collapse into a black hole without an explosion \citep{Heger03}.
\end{itemize}

The hydrogen ionising radiation emitted during main sequence are
propagated with adaptive ray tracing \citep{Abel02_RT, Wise11_Moray}
that is based on the HEALPix framework \citep{HEALPix}.  The radiation
field is evolved at every hydrodynamics timestep of the finest AMR
level that is on the order of 10 kyr.  The photo-heating that occurs
on these timesteps couples the radiation effects to the hydrodynamics.
In addition, this simulation considers the momentum transfer from
ionising radiation to the absorbing gas, i.e. radiation pressure.  We
model the LW radiation with a time-dependent optically-thin soft UV
background and add inverse square profiles, centred on all Pop II and
III star particles, to estimate the total LW radiation at all points.
Using this physics set, we have shown in W12 that the galaxies
produced in this simulation match the luminosity-metallicity relation
for local dwarf galaxies and does not suffer from the well-known
galaxy overcooling problem seen in many previous galaxy formation
simulations.

\section{Simulation Results}
\label{sec:results}

\begin{figure}
  \centering
  \includegraphics[width=\columnwidth]{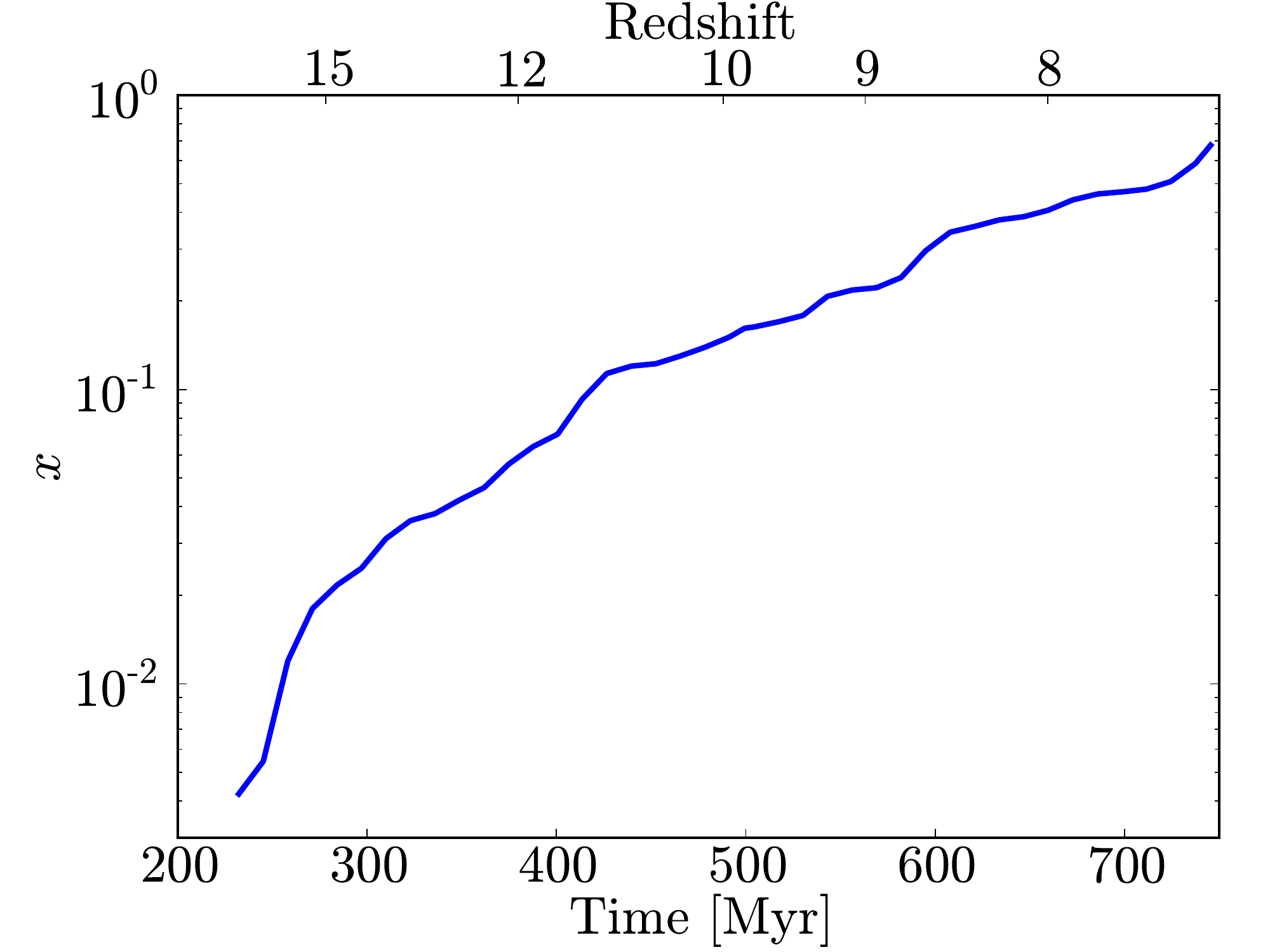}
  \caption{Ionisation history of the simulation, showing the ionised
    volume fraction $x$.  We consider a computational element to be
    ionised if $x \ge 0.5$.  Reionisation in the simulation starts
    when the first Pop III stars form in the calculation at $z = 17$.}
  \label{fig:ion}
\end{figure}

The previous two papers in this series studied the individual
properties of selected galaxies\footnote{We define a galaxy as any
  metal-enriched stellar system that is gravitationally bound and
  exists in a dark matter halo \citep{Willman12}.}.  In this work, we
study the global statistics of galaxies, in particular, their stellar
masses, SFRs, gas fractions, LF, the fraction of escaping UV
radiation, and their contribution to reionisation.  The stellar
radiation from these galaxies reionise 70 per cent of the simulation
volume by $z=7.3$, and the ionisation history is shown in Figure
\ref{fig:ion}.  It should be noted that this history is not
cosmologically representative as a simulation box of $\ga 100\,
\textrm{Mpc/h}$ is required \citep[e.g.][]{Iliev06, Iliev13} for such
a study.  However the SFRs and UV escape fractions are important to
quantify in small volume, high-resolution simulations, so they can
guide radiation source models in large volume reionisation
calculations needed for predictions for a cosmic reionisation history.

\subsection{Radiative cooling in low-mass haloes}
\label{sec:cooling}

\begin{figure}
  \centering
  \includegraphics[width=\columnwidth]{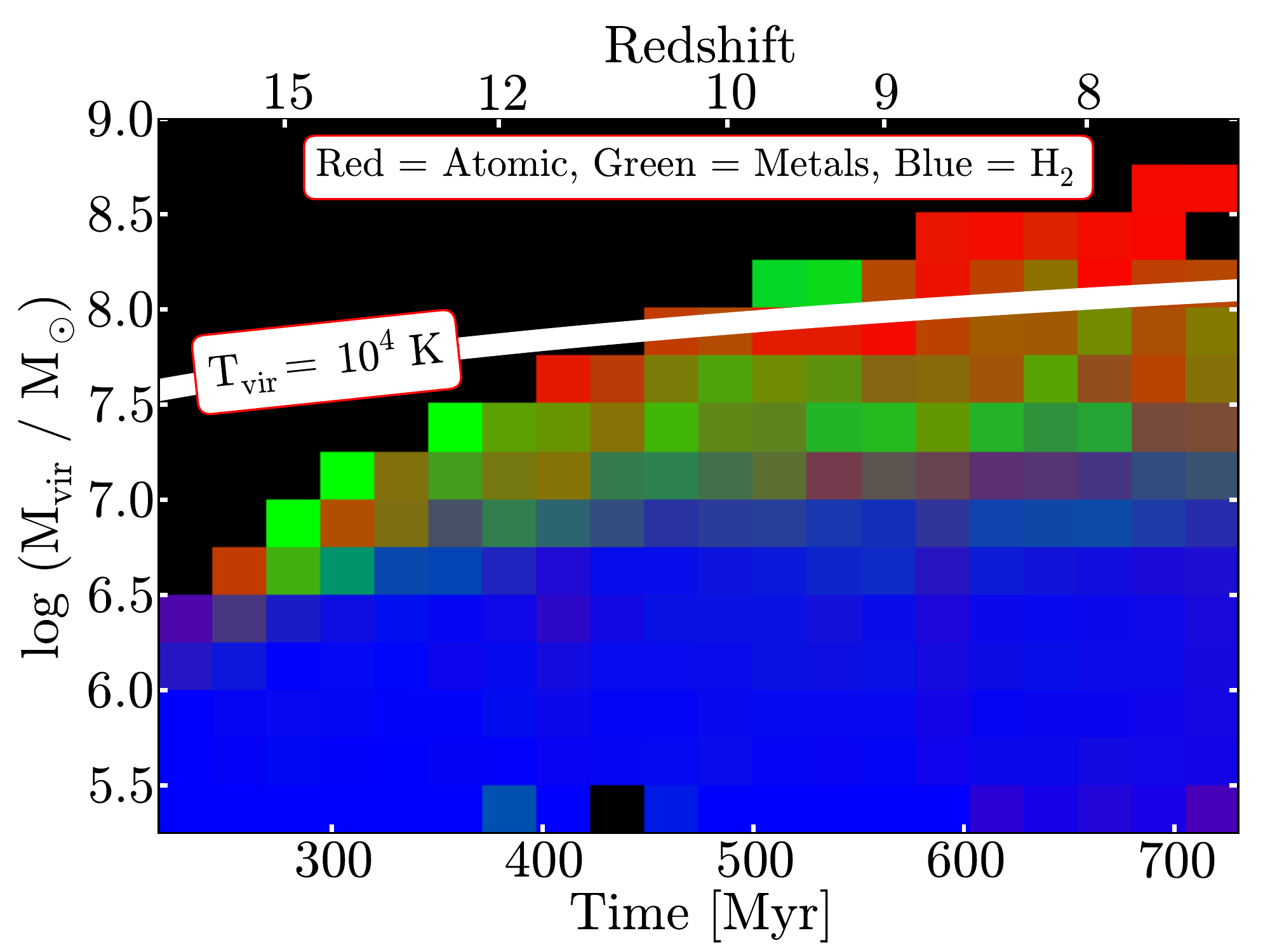}
  \caption{Fractional radiative cooling rates from atomic processes
    (red), fine-structure metal transitions (green), and
    \hh~transitions (blue) as a function of time and halo mass,
    displayed as \textit{RGB} colours and averaged over all haloes.
    The sum of each \textit{RGB} channel in a cell is normalized to
    unity.  The thick white line shows the halo mass corresponding to
    a virial temperature of $10^4 \unit{K}$.  There is a clear
    transition from \hh-cooling to metal-line cooling at $M_{\rm vir}
    \ga 10^7 \Ms$ and to atomic cooling at $T_{\rm vir} = 10^4
    \unit{K}$.  Halos with $T_{\rm vir} < 10^4 \unit{K}$ can partially
    cool through recombination radiation in \hii~regions, depicted by
    the olive and purple shades for metal-enriched and metal-free
    halos, respectively.}
  \label{fig:hcool}
\end{figure}
\begin{figure}
  \centering
  \includegraphics[width=\columnwidth]{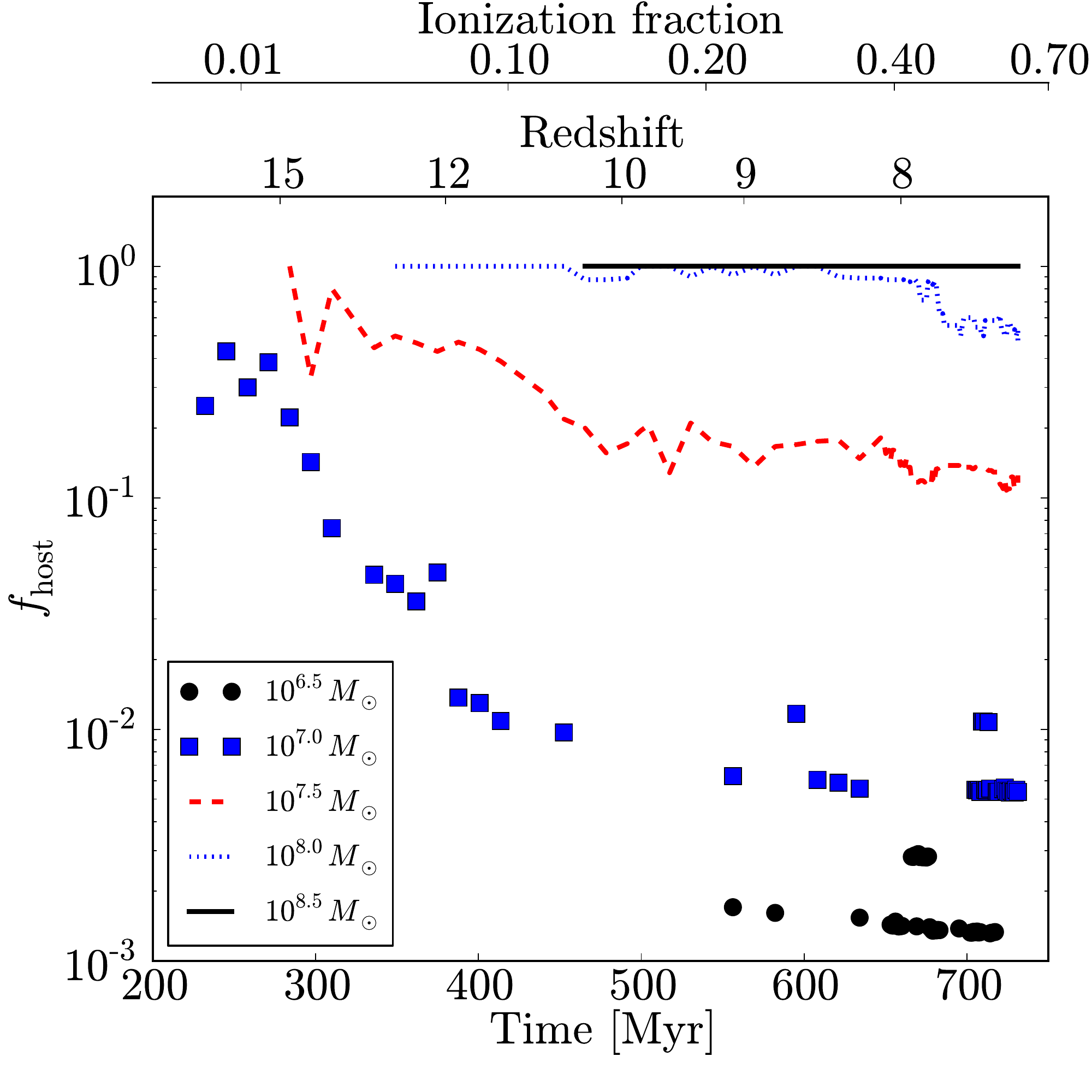}
  \caption{The fraction of haloes that host metal-enriched stars as a
    function of time in five different mass bins.  The corresponding
    ionization fraction is shown on the upper axis.  See Figure
    \ref{fig:ion} for a plot of ionization history.  Star formation in
    the non-atomic cooling haloes ($\mvir < 10^8 \Ms$) occurs at
    high-redshift, only to be suppressed by external radiative
    feedback.}
  \label{fig:occupy}
\end{figure}

It has generally been thought that galaxies begin to form in haloes
that can support hydrogen atomic cooling with $T_{\rm vir} \ga 10^4$~K
because molecular hydrogen is easily dissociated when the halo is
exposed to a UV radiation field.  However, previous numerical
studies have shown that \hh~formation and radiative cooling occurs in
lower mass haloes, even in the presence of a strong LW
radiation field \citep{Wise07_UVB, OShea08}.  In addition if
the halo is chemically enriched, then metal-line radiative cooling
adds to the likelihood that the halo will collapse and form stars.

Figure \ref{fig:hcool} illustrates these additional cooling processes
in low-mass haloes, where we show the average fractional radiative
cooling rates in haloes as a function of time and mass.  For each
halo, we calculate the total cooling rate $\Lambda_{\rm tot}$ within
the virial radius from hydrogen and helium processes $\Lambda_{\rm X}$
using the rates of \citet{Sutherland93}, metal cooling $\Lambda_{\rm
  Z}$ using the method of \citet{Smith08_Cooling}, and molecular
hydrogen cooling $\Lambda_{{\rm H}_2}$ using the rates of
\citet{Glover08_Rates}.  We depict the fractional cooling rate $f_{\rm
  cool,i} \equiv \Lambda_i/\Lambda_{\rm tot}$, where $i =$ (X,Z,\hh),
in a composite \textit{RGB} image with each \textit{RGB} channel
having a value equal to $255f_{\rm cool,i}$.  We weight the haloes by
the time between outputs (see \S\ref{sec:sim}) when calculating the
probability distribution function (PDF).

Three distinct modes of cooling in these low-mass haloes are apparent
in this analysis.  The lowest mass haloes cool by \hh-transitions
because they are still metal-free and not massive enough to sustain
atomic cooling.  However in halos with $M \ga 10^7 \Ms$, Pop III SNe
occurring in progenitor or nearby haloes enrich the gas, resulting in
metal-line cooling being dominant.  In this work, we will show that a
non-negligible amount of star formation occurs in such haloes, which
we term metal-cooling (MC) haloes hereafter.  In halos with $T_{\rm
  vir} \ga 10^4 \unit{K}$, atomic radiative processes are the main
coolant.  Furthermore, recombinations in \hii~regions provide
additional cooling in haloes below the atomic cooling limit that have
formed stars, depicted by the olive and purple shades in the PDF.

\begin{table*}
  \caption{General galaxy and host halo properties}
  \label{tab:stats}
  \centering
  \begin{tabular*}{0.9\textwidth}{@{\extracolsep{\fill}}c c c c c c c c}
    \hline\hline\\[-2.5ex]
    $\log M_{\rm vir}$ & $\log M_\star$ & $\langle f_{\rm host}
    \rangle$ & $\log f_\star$ &
    $f_{\rm gas}$ & M$_{\rm UV}$ & $\log L_{\rm bol}$ & $\log \dot{N}_{\rm ion}$\\
    $[M_\odot]$ & $[M_\odot]$ & & & & & $[L_\odot]$ &
    $[\textrm{ph s}^{-1}]$\\
    (1) & (2) & (3) & (4) & (5) & (6) & (7) & (8)\\[1ex]
    \hline\\[-1.5ex]
    6.5 & $3.41^{+0.10}_{-0.06}$ & $2.4 \times 10^{-4}$ & $-1.88^{+0.35}_{-0.43}$ &
    $0.05^{+0.02}_{-0.03}$ & $-5.61^{+0.73}_{-0.86}$ &
    $4.81^{+0.20}_{-0.31}$ & $49.2^{+0.1}_{-0.2}$\\[1ex]
    7.0 & $3.59^{+0.24}_{-0.33}$ & 0.052 & $-2.25^{+0.11}_{-0.50}$ &
    $0.06^{+0.02}_{-0.03}$ & $-6.09^{+1.35}_{-0.93}$ &
    $5.02^{+0.26}_{-0.55}$ & $49.5^{+0.2}_{-0.1}$\\[1ex]
    7.5 & $3.88^{+0.18}_{-0.29}$ & 0.28 & $-2.48^{+0.27}_{-0.32}$ &
    $0.07^{+0.01}_{-0.02}$ & $-7.17^{+1.30}_{-1.22}$ &
    $5.43^{+0.42}_{-0.50}$ & $49.6^{+0.3}_{-0.4}$\\[1ex]
    8.0 & $4.60^{+0.30}_{-0.12}$ & 0.90 & $-2.25^{+0.15}_{-0.04}$ &
    $0.10^{+0.02}_{-0.02}$ & $-9.59^{+0.87}_{-0.95}$ &
    $6.39^{+0.39}_{-0.39}$ & $49.9^{+0.3}_{-0.3}$\\[1ex]
    8.5 & $5.74^{+0.31}_{-0.37}$ & 1.0 & $-1.80^{+0.23}_{-0.21}$ &
    $0.13^{+0.00}_{-0.00}$ & $-13.72^{+0.82}_{-0.72}$ &
    $8.06^{+0.26}_{-0.31}$ & $51.6^{+0.1}_{-0.1}$\\[1ex]
    \hline
  \end{tabular*}
  \parbox[t]{0.9\textwidth}{\textit{Notes:} Statistics are shown for
    galaxies at all times in 0.5 dex bins in \mvir.  Column (1): Virial
    mass.  Column (2): Stellar mass.  Column (3): Time-averaged galaxy
    occupation fraction of halos.  See Figure \ref{fig:occupy} for the
    time dependent of this fraction.  Column (4): Star formation efficiency,
    $M_\star / M_{\rm gas}$.  Column (5): Gas mass fraction.
    Column (6): UV magnitude at 1500 \AA.  Column (7): Bolometric
    luminosity.  Column (8): Ionizing photon luminosity from young ($<20$
    Myr) stars.  Errors shown are 1-$\sigma$ deviations.}
\end{table*}



\subsection{Dwarf galaxy properties}
\label{sec:stats}

At the final redshift $z = 7.3$ of our simulation, there are 32 haloes
that host metal-enriched star formation.  The smallest such halo has a
virial mass $M_{\rm vir} = 7.6 \times 10^5 \Ms$ and stellar mass
$M_\star = 840 \Ms$, which has been enriched by a nearby SN and did
not form any Population III stars.  The largest halo has $M_{\rm vir}
= 6.8 \times 10^8 \Ms$ and $M_\star = 3.7 \times 10^6 \Ms$.  Here we
define the halo as a sphere that contains an overdensity of $18\pi^2$
relative to the critical density.  There exists a wide range of
stellar and gaseous mass fractions between these extremes.  Because
these low-mass haloes are heavily affected by stellar feedback, they
cycle between quiescence and star-forming phases during their
histories \citep[e.g.][]{Wise12_Galaxy, Hopkins13_FIRE}.  To capture
the whole range of halo and galaxy properties from our small set of
simulated dwarf galaxies, we present our results from \textit{all
  redshifts} until the final redshift.  During inhomogeneous
reionization, we have found that MC halo properties are more dependent
on environment (i.e. neutral or ionised) instead of halo mass because
they must re-accrete gas to form stars after Pop III stellar feedback
ejects the majority of their gas.  Figure \ref{fig:occupy} illustrates
the evolution of galaxy occupation fraction $f_{\rm host}$,
i.e. haloes that host metal-enriched stars, that is divided into five
different mass bins with a width of 0.5 dex.  The time-averaged values
of $f_{\rm host}$ are shown in Table \ref{tab:stats}.  Because the
galaxy formation suppression shown in Figure \ref{fig:occupy} is
dependent on the particular reionization history of this simulation,
the ionization fraction $x$, not the exact redshifts in this small box
simulation, should be used when applying these data.

\begin{figure*}
  \centering
  \includegraphics[width=\textwidth]{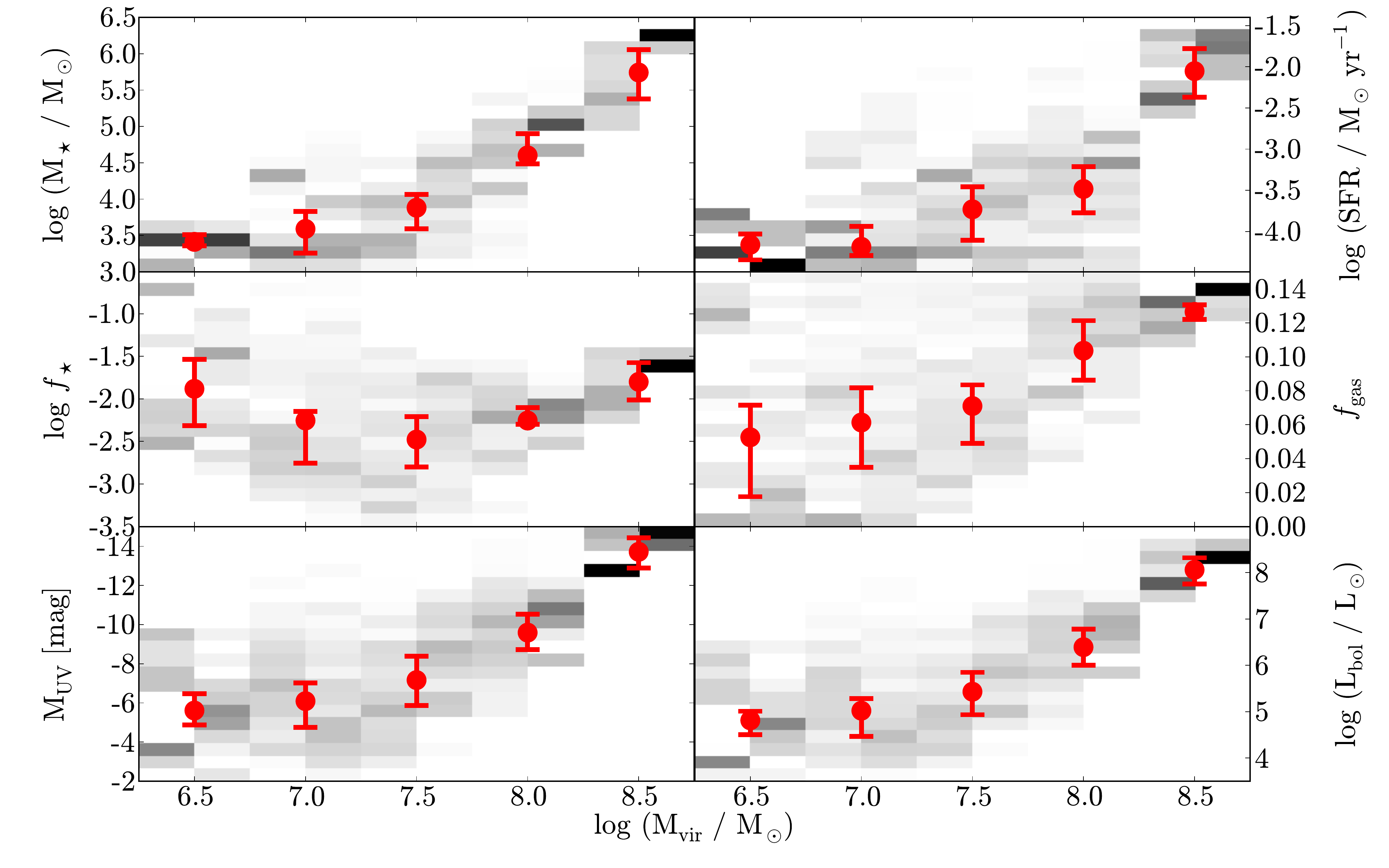}
  \caption{(\textit{Clockwise from the upper left}) Stellar mass, star
    formation rate, gas mass fraction, bolometric stellar luminosity,
    absolute UV magnitude at 1500~\AA~from the stellar component, and
    star formation efficiency of all star-forming haloes at all
    redshifts.  The pixels show the probability distribution function
    that is weighted by the time between snapshots.  The points and
    error bars show the mean values and standard deviations in 0.5 dex
    bins, respectively.  The increase in SFRs and luminosities at
    $10^8 \Ms$ are caused by higher gas fractions and more efficient
    radiative cooling through atomic hydrogen line transitions.  Below
    this mass scale, star formation is induced primarily through metal
    fine-structure line and \hh~cooling.}
  \label{fig:stats}
\end{figure*}

At the low-mass range at early times, galaxies form in about one-third
of the haloes with $\mvir = 10^7 \Ms$, and this fraction steeply
decreases after $z \sim 15$ ($x \sim 0.02$) to one per cent within a
halo sound crossing time $t_{\rm sc} \sim 100 \unit{Myr}$
\citep{Shapiro04}.  These low-mass galaxies are some of the first to
form in the simulation and ionise the surrounding IGM.  The subsequent
suppression occurs because the haloes must then accrete from a
pre-heated IGM.  The filtering mass,
\begin{equation}
  \label{eqn:filter}
  M_{\rm F}^{2/3} (a) = \frac{3}{a} \int_0^a da^\prime \,
  M_{\rm J}^{2/3}(a^\prime) \left( 1 - \frac{a^\prime}{a} \right),
\end{equation}
is an analytical measure of the minimum mass of a Jeans-unstable halo
in such a situation.  Here $a$ and $M_{\rm J}$ are the scale factor
and time-dependent Jeans mass of the accreted gas \citep{Hui98,
  Gnedin00}.  The Jeans mass in the ionised regions is approximately
$4 \times 10^8 \Ms$ at $z = 20$ and increases to $10^9 \Ms$ by $z = 7$
for a gas at the mean baryon density and $T = 10^4$~K.  The filtering
mass correspondingly increases from $10^7 \Ms$ at $z = 15$ to $10^8
\Ms$ at $z = 7$ \citep[see][]{Wise12_Galaxy}.  Similar but not as
severe suppression occurs at $\mvir = 10^{7.5} \Ms$ in which $f_{\rm
  host}$ is around 0.5 at $z \ga 12$ ($x \sim 0.1$) and decreases to
$\sim 0.15$ afterwards.  This remaining 15 per cent of galaxies are
not actively forming stars, but they host stars that formed in their
initial star formation event and have not accreted additional mass.
Atomic cooling haloes with $\mvir \ge 10^8 \Ms$ all host galaxies at $z
\ge 8$.  After this time, star formation in half of the $\mvir = 10^8
\Ms$ haloes is suppressed because the filtering mass is now $\sim 10^8
\Ms$.

Although mass accretion and star formation histories differ from halo
to halo, the stellar and gaseous properties exhibit general trends
with respect to halo mass with some inter-halo variations.  Figure
\ref{fig:stats} shows the time-averaged stellar masses $M_\star$,
SFRs, star formation efficiencies $f_\star = M_\star / M_{\rm gas}$,
gas fraction $f_{\rm gas} = M_{\rm gas} / M_{\rm vir}$, bolometric
luminosity $L_{\rm bol}$, and AB magnitude $M_{\rm UV}$ at 1500\AA~of
all of the galaxies as a function of halo mass.  We calculate $L_{\rm
  bol}$ and $M_{\rm UV}$ using the stellar population synthesis model
\textsc{galaxev} \citep{Bruzual03}, using the star particle ages,
masses, and metallicities as input into \textsc{galaxev} and assuming
an instantaneous burst model for each Pop II star particle.  We do not
consider any nebular emission.  The shaded regions in the figure show
the PDF that is weighted by the time between outputs and is normalised
in each column.  The points show the average values and 1-$\sigma$
deviations in bins of 0.5 dex, which are also shown in Table
\ref{tab:stats}.

The stellar mass and SFRs in MC haloes exhibit less of a dependence on
halo mass than atomic cooling haloes with $M_\star$ increasing from
$10^{3.4}$ to $10^{3.9} \Ms$ over a halo mass range $\log \mvir =
[6.5, 7.5]$.  In comparison, $M_\star$ increases by almost two orders
of magnitude over the next decade of halo mass.  Correspondingly,
similar behaviour is seen in their SFRs, absolute UV magnitudes, and
bolometric luminosities.  What causes this change in behaviour?  We
can gain some insight by exploring how $f_{\rm gas}$ and $f_\star$
vary with halo mass.  We see that the average value of $f_\star$ only
varies by a factor of three around 0.01 over the entire mass range
with a minimum at $\mvir = 10^{7.5} \Ms$.  This concavity in $f_\star$
is caused after the galaxy experiences its first Pop II star formation
event, whose feedback suppresses any further star formation until the
gas reservoir is replenished.  During this suppression phase and after
the outflows have ceased, the halo gas mass increases as the stellar
mass remains constant, leading to a decrease in $f_\star$ as shown
in the plot.  Furthermore, once the halo can cool through atomic line
transitions, the star formation efficiency increases.  On the other
hand, the average value of $f_{\rm gas}$ steadily increases from 0.05
at $\mvir = 10^{6.5} \Ms$ up to 0.13 in $10^{8.5} \Ms$ haloes.  The
larger scatter in low mass haloes originates from their susceptibility
to internal and external radiative and SN feedback.  Varying degrees
of feedback are caused by differing Pop III stellar masses that their
progenitors hosted.  At $\mvir > 10^8 \Ms$, the gravitational
potential well is deep enough so that the outflows only contain a
small mass fraction.  Thus, we conclude that both the increase in the
star formation efficiency and gas mass fraction cause the greater star
formation rates as haloes transition from \hh~and metal-line cooling
to atomic line cooling.

\subsection{Luminosity functions}

We have shown that galaxies begin to form in haloes that rely on
\hh~and metal-line cooling but not atomic line cooling.  Recall that
these haloes can have masses below the filtering mass.  During this
phase, star formation is less efficient than larger haloes with $T \ga
10^4$~K.  Afterward, they transition to more efficient star formation
in atomic cooling haloes, which can be gradual or sudden if a galaxy
experiences a major merger or not, respectively.  These two different
modes of galaxy formation should manifest in the galaxy luminosity
function at high ($M_{\rm UV} \ga -15$) magnitudes.

\begin{figure}
  \centering
  \includegraphics[width=\columnwidth]{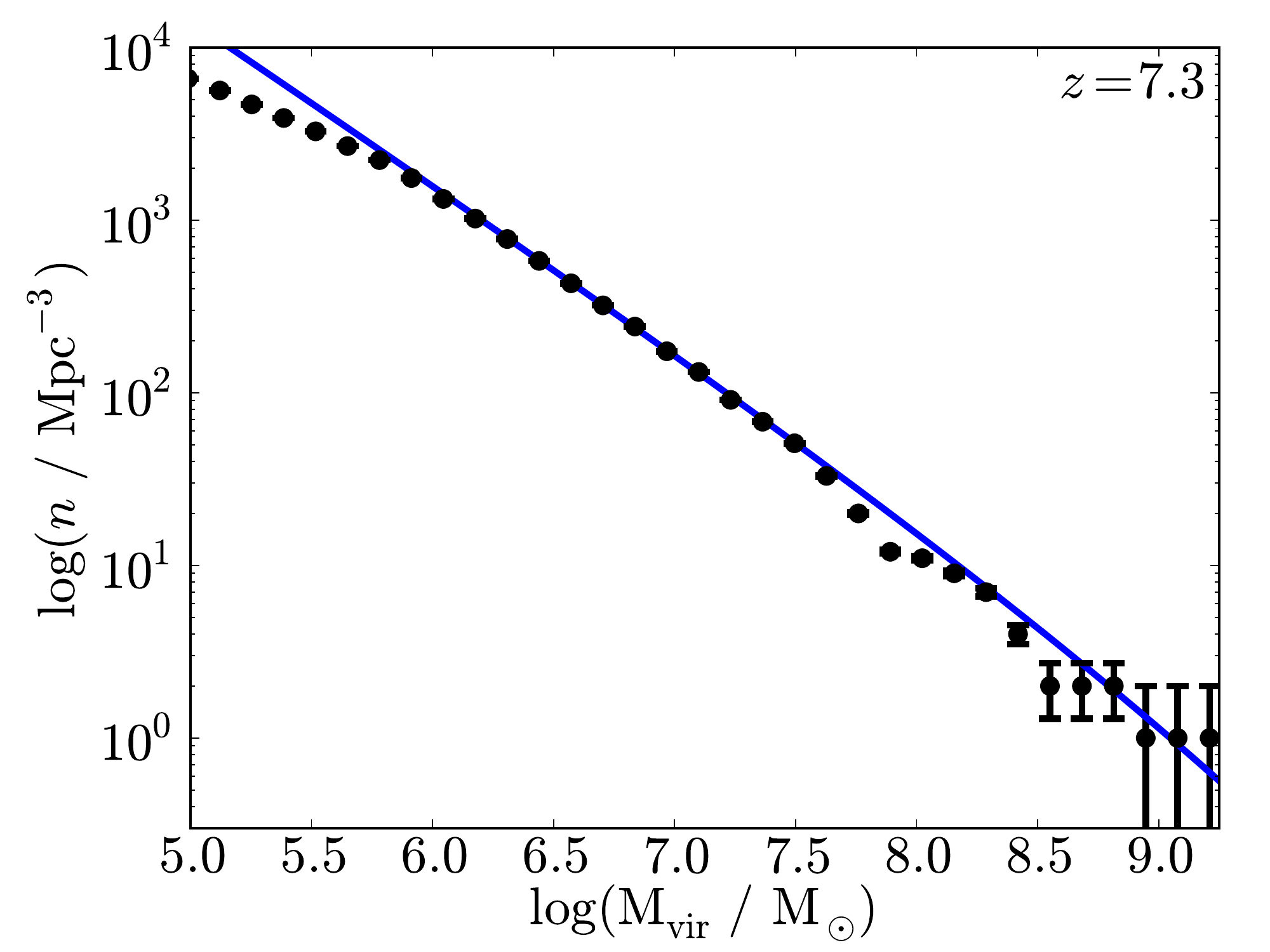}
  \caption{Halo mass function from the simulation (circles) and the
    analytical fit from \citet{Warren06}.  The error bars depict
    Poisson noise.}
  \label{fig:massfn}
\end{figure}

Before constructing LFs from our data, we first check whether our
small-scale (1 comoving Mpc$^3$) simulation is representative of the
cosmic mean.  Figure \ref{fig:massfn} shows the simulated halo mass
function against the analytical mass function of \citet{Warren06} that
is calibrated against a suite of N-body simulations.  In our simulated
volume, the halo mass function is consistent with analytical
expectations at $M > 5 \times 10^5 \Ms$, and below this mass scale,
the simulated halo mass function is below the analytical fit because
these small haloes are not well resolved, which does not affect the
luminosity function because galaxies do not form in such small haloes.

\begin{table}
  \caption{Mean dwarf galaxy luminosity function at $M_{\rm UV} > -12$}
  \label{tab:lf}
  \centering
  \begin{tabular*}{0.5\columnwidth}{@{\extracolsep{\fill}}c c}
    \hline\hline
    Redshift & $\phi\; [\mathrm{mag}^{-1} \mathrm{Mpc}^{-3}]$\\
    \hline
    7.3 & $2.2 \pm 1.4$\\
    8 & $2.3 \pm 1.2$\\
    9 & $2.5 \pm 1.2$\\
    10 & $1.5 \pm 0.8$\\
    12 & $1.2 \pm 0.6$\\
    15 & $0.7 \pm 0.2$\\
    \hline
  \end{tabular*}
\end{table}



\begin{figure*}
  \centering
  \includegraphics[width=\textwidth]{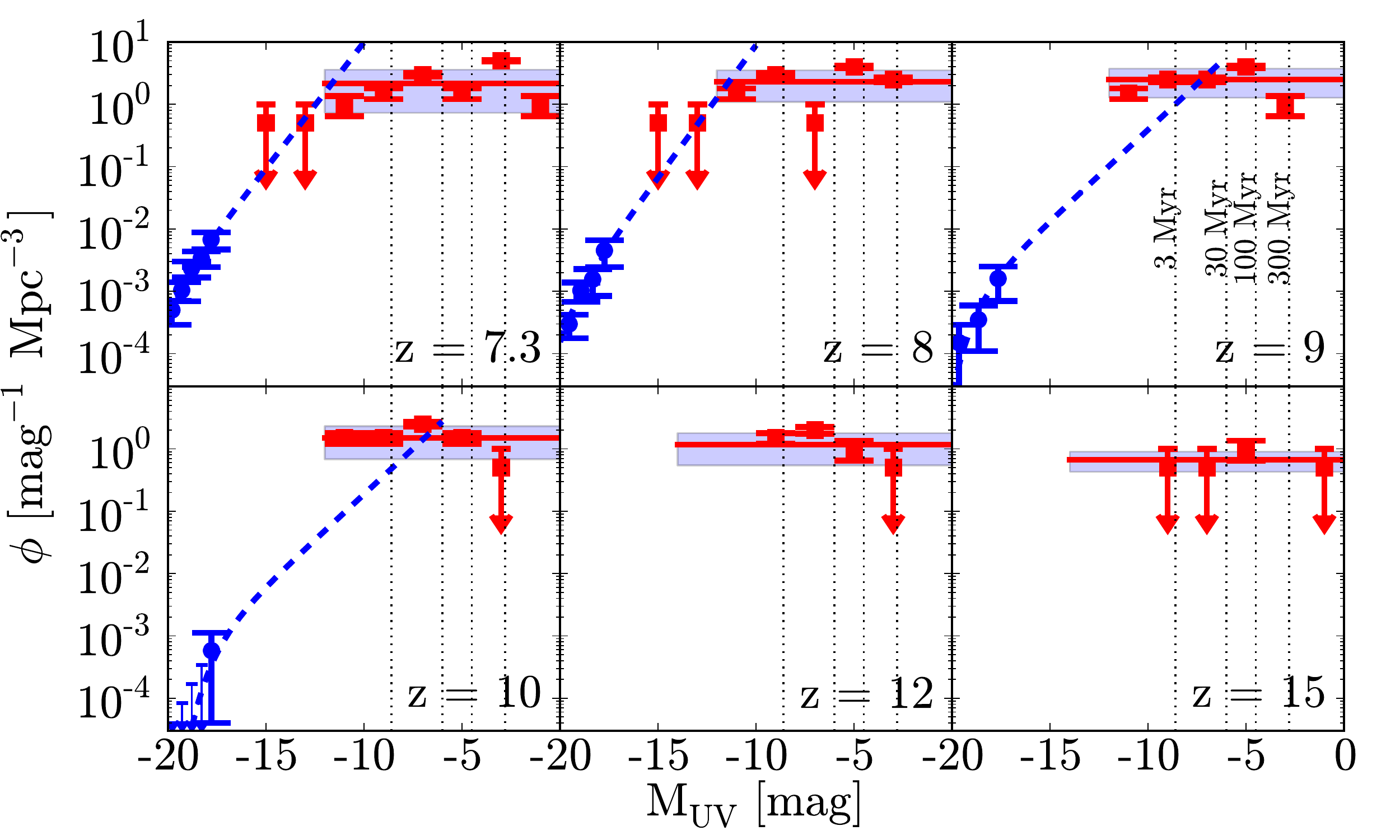}
  \caption{Galaxy LFs from the simulation (red squares) and the Hubble
    Ultra Deep Field (blue circles) at redshifts 7, 8, 9, 10, 12, and
    15.  The blue dashed line shows the fits from \citet{Bouwens11}
    for the redshift 7 and 8 data and \citet{Oesch13} for the redshift
    9 and 10 data.  The error bars depict Poisson noise, and the
    points with arrows represent data where only one galaxy exists in
    bin.  The red solid line and surrounding shaded region shows the
    average number density and standard deviation of galaxies with
    $M_{\rm UV} > -12$.  The vertical dotted lines show the UV
    magnitude of one star particle at our resolution limit of $M_\star
    = 1000 \Ms$ at various ages.}
  \label{fig:lf}
\end{figure*}

In order to understand the behaviour of the galaxy LF at $M_{\rm UV}
\ga -15$, we calculate the simulated LF at several redshifts $z =
[7.3, 8, 9, 10, 12, 15]$, which are shown in Figure \ref{fig:lf} in
red squares.  For comparison, we also plot the observed galaxy LFs and
their fits from the HUDF galaxies at $z \le 10$ \citep{Bouwens11,
  Oesch13}.  It is clear that these dwarf galaxies do not follow the
same power law found in $L^*$ galaxies at similar redshifts, but their
LF is nearly flat at $M_{\rm UV} \ga -12$.  The mean values and their
standard deviations are displayed in Table \ref{tab:lf}.  All of the
galaxies with these magnitudes exist in MC haloes with $\mvir \le 10^8
\Ms$ (see Figure \ref{fig:stats}).  The comoving density of these
haloes only varies by a factor of a few during the redshift range
7--15 \citep{Mo02}.  Effects from radiative and SN feedback discussed
in the previous section and the weakly varying halo number density can
explain this flat LF at $M_{\rm UV} \ga -12$.  At a sufficiently high
absolute magnitude, the galaxy LF should start to approach zero;
however, our stellar mass resolution does not allow us to answer this
question.  In Figure \ref{fig:lf}, we have marked the $M_{\rm UV}$ of
a single 1000~\Ms~stellar cluster with various ages, showing the
stellar mass resolution limit.  We note that the simulation contains a
few star particles below 1000~\Ms~(see \S\ref{sec:sf} for an
explanation), which result in the dimmest galaxies at high-redshift.

Only at $z \le 8$ in the simulation, a few galaxies with $M \la -12$
are massive enough to host efficient star formation in atomic cooling
haloes, and their number densities are consistent with the $z=7$ and
$z=8$ HUDF LFs.  As seen in Figure \ref{fig:stats}, nearly all of the
galaxies with $M_{\rm UV} \le -12$ are in the $\mvir = 10^{8.5} \Ms$
mass bin, and the galaxies dimmer than this threshold are hosted by
less massive haloes.  This kink in the LF signifies the transition
from \hh~and metal-line cooling to atomic cooling.  The intersection
between the LFs from the HUDF galaxies and our simulated galaxies is
highly dependent on the slope $\alpha$.  In the $z=9$ and $z=10$
observed LF, the value for $\alpha$ is fixed to $-1.73$
\citep{Oesch13}, whereas at $z=7$ and $z=8$, there exists enough
sub-L$^*$ galaxies to fit $\alpha = -2.01 \pm 0.21$ and $-1.91 \pm
0.32$, respectively.  Because we have found the LF to be flat and
time-independent at $M_{\rm UV} \ga -12$, the transition to a flat LF
should only dependent on the normalization $\phi^*$ and slope $\alpha$
of more luminous galaxies.

\subsection{Escape fraction}

The progression of cosmic reionization from stellar sources is
inherently dependent on the ionizing photon luminosity, which
primarily originate from massive stars, and how much of this radiation
escapes into the IGM.  The production rate of ionizing photons
$\dot{n}_{\rm halo, \gamma}$ that escape from a particular halo can be
parameterised into a product of four efficiency factors,
\begin{equation}
  \label{eqn:ngamma_halo}
  \dot{n}_{\rm halo, \gamma} = f_{\rm esc} f_\gamma f_{\rm gas}
  f_\star M_{\rm vir}.
\end{equation}
Here the total stellar mass $M_\star = f_{\rm gas} f_\star M_{\rm
  vir}$, and $f_\gamma$ is the number of ionizing photons produced per
stellar baryon, which can range from 6000 for a Salpeter IMF with
solar metallicity to 13,000 for a metal-poor ([Z/H] = --3.3)
population \citep{Schaerer03}.  Then a fraction $f_{\rm esc}$ of the
emitted photons escape into the IGM.  This quantity is the most
uncertain of the four factors that enter into reionization
calculations and has been the focus of several observational and
numerical campaigns.

\subsubsection{Method}

Radiative and SN feedback can create channels of diffuse
ionised gas from the galaxy centre to the IGM, where photons can
escape from the halo relatively unobscured.  Because our simulation
includes both of these effects, we can calculate the UV escape
fraction in post-processing without any loss in accuracy.  However, we
do lose any variations that might occur between data outputs.

We define $f_{\rm esc}$ as the fraction of stellar photons that exit a
sphere with radius \rvir~located at the halo center of mass.  The UV
escape fraction is directly related to the \hi~column density
\nhi~between each star and this sphere.  We first divide the spherical
surface into 768 HEALPix (level 3) equal-area pixels. Then for each
star particle, we compute \nhi~between the particle and the $i$-th
pixel, which is converted into an transmittance $T_i$ in that
particular line of sight,
\begin{equation}
  T_i = \exp(-\sigma_{\rm HI} N_{{\rm HI},i}),
\end{equation}
where we take the \hi~cross-section $\sigma_{\rm HI} = 1.78 \times
10^{-18} \unit{cm}^2$ at 21.6~eV, which is the same energy as the
radiation considered in the simulation \citep{Verner96}.  Because the
sphere is divided into equal-area pixels, the escape fraction for a
single star particle is the average transmittance $\overline{T}$ over
all of the pixels, and $f_{\rm esc}$ for the entire halo is the
luminosity-weighted average of $\overline{T}$,
\begin{equation}
  f_{\rm esc} = \sum_{n}^{\rm stars} L_{{\rm ion},n}
    \overline{T}_n \Big/ \sum_{n}^{\rm stars} L_{{\rm ion},n},
\end{equation}
where $L_{{\rm ion},n}$ is the ionizing luminosity in the halo.

\subsubsection{Global trends}

\begin{table*}
  \caption{UV escape fraction}
  \label{tab:fesc}
  \centering
  \begin{tabular*}{0.9\textwidth}{@{\extracolsep{\fill}}c c c c c c c}
    \hline
    & \multicolumn{3}{c}{Mean} &
    \multicolumn{3}{c}{Luminosity-weighted mean}\\
    \cline{2-4} \cline{5-7}\\[-2ex]
    $\log M_{\rm vir}$ & $f_{\rm esc}$ & $\log f_{\rm esc} f_\star$ &
    $\log f_{\rm esc} f_\star f_{\rm gas}$ & $f_{\rm esc}$ & $\log f_{\rm esc} f_\star$ &
    $\log f_{\rm esc} f_\star f_{\rm gas}$\\
    (1) & (2) & (3) & (4) & (5) & (6) & (7)\\[1ex]
    \hline\\[-1.5ex]
    6.5 & $0.49^{+0.21}_{-0.26}$ & $-2.75^{+0.15}_{-0.37}$ &
    $-3.91^{+0.22}_{-0.10}$ & $0.52^{+0.24}_{-0.29}$ &
    $-2.44^{+0.25}_{-0.27}$ & $-3.60^{+0.09}_{-0.32}$\\[1ex]
    7.0 & $0.45^{+0.18}_{-0.21}$ & $-2.80^{+0.53}_{-0.60}$ &
    $-3.92^{+0.52}_{-0.52}$ & $0.54^{+0.13}_{-0.24}$ &
    $-2.26^{+0.50}_{-0.18}$ & $-3.21^{+0.52}_{-0.10}$\\[1ex]
    7.5 & $0.27^{+0.14}_{-0.20}$ & $-3.25^{+0.40}_{-0.29}$ &
    $-4.36^{+0.46}_{-0.23}$ & $0.32^{+0.16}_{-0.16}$ &
    $-2.60^{+0.35}_{-0.29}$ & $-3.52^{+0.37}_{-0.40}$\\[1ex]
    8.0 & $0.10^{+0.01}_{-0.08}$ & $-3.57^{+0.36}_{-0.29}$ &
    $-4.54^{+0.32}_{-0.32}$ & $0.25^{+0.11}_{-0.19}$ &
    $-2.90^{+0.67}_{-0.50}$ & $-3.79^{+0.68}_{-0.58}$\\[1ex]
    8.5 & $0.04^{+0.01}_{-0.03}$ & $-3.58^{+0.59}_{-0.60}$ &
    $-4.48^{+0.61}_{-0.60}$ & $0.05^{+0.01}_{-0.02}$ &
    $-3.00^{+0.27}_{-0.05}$ & $-3.88^{+0.27}_{-0.04}$\\[1ex]
    \hline
  \end{tabular*}
  \parbox[t]{0.9\textwidth}{\textit{Notes:} 
    Statistics are shown for galaxies at all times in 0.5 dex bins in
    \mvir.  Column (1): Virial mass in units of \Ms.  Column (2):
    Fraction of hydrogen ionizing radiation that escapes the virial
    radius.  Column (3): Product of the UV escape fraction and star
    foramtion efficiency. Column (4): Product of column (3) and the
    gas mass fraction. Columns (5)-(7): Same as Columns (2)-(4) but
    ionizing luminosity weighted means are shown. Errors shown are
    1-$\sigma$ deviations.
  }
\end{table*}



\begin{figure}
  \centering
  \includegraphics[width=\columnwidth]{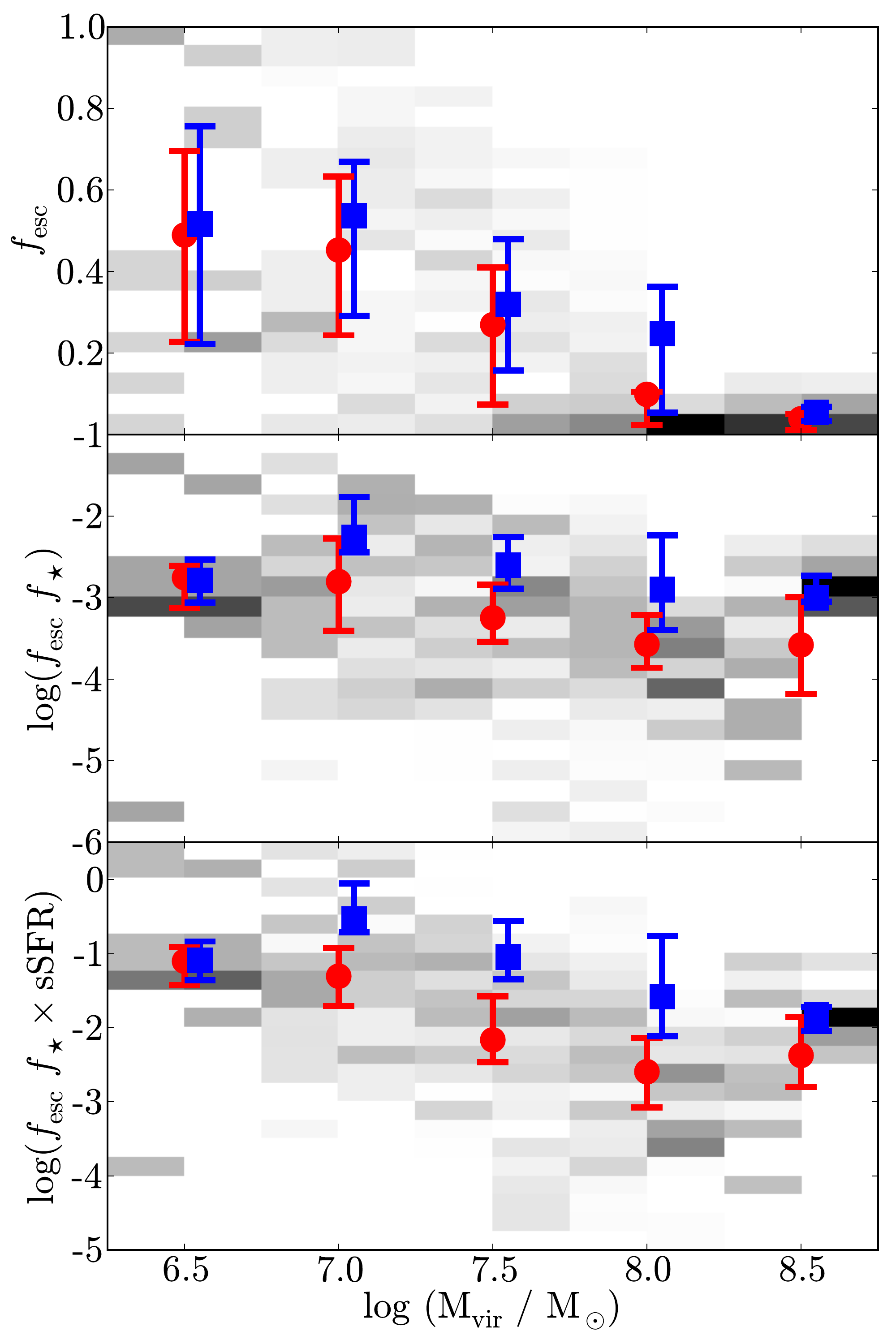}
  \caption{Probability density functions (shaded regions) for the UV
    escape fraction $f_{\rm esc}$ (top panel), product of the star
    formation efficiency $f_\star = M_\star / M_{\rm gas}$ and escape
    fraction (middle panel), and product of the middle panel and the
    gas mass fraction (bottom panel) as a function of halo mass.  The
    data are shown from all times, and the pixels are weighted by the
    time between snapshots.  The mean and luminosity-weighted mean
    values in 0.5 dex bins are represented by red circles and blue
    squares, respectively.  The blue squares have been offset to the
    right for clarity.}
  \label{fig:fesc}
\end{figure}

Our main results for the escape fraction are shown in Figure
\ref{fig:fesc}.  We analyze the data similar to our approach with the
average halo properties, where we calculate the PDF for all {\it
  actively star-forming} haloes at all times.  Here we only include
galaxies with young ($< 20$~Myr) stars because massive OB stars emit
almost all of the ionizing photons in a given stellar population.  In
addition, we use this same age criterion to determine which star
particles emit ionizing radiation in the simulation.  The red circles
in this Figure represent the mean quantity in 0.5 dex bins of \mvir,
and the blue squares denote the luminosity-weighted average of the
quantities.  These averages and their 1-$\sigma$ deviations are listed
in Table \ref{tab:fesc}.

The top panel of Figure \ref{fig:fesc} shows the behaviour of
\fesc~with respect to halo mass.  The lowest mass haloes with masses
$10^{6.25} - 10^{7.25} \Ms$ have \fesc~around 50 per cent for both the
mean and luminosity-weighted averages.  This fraction then decreases
with increasing halo mass.  In the $10^{7.5} \Ms$ bin, \fesc~is 30 per
cent, shrinking to a mean and luminosity-averaged value of 10 and 25
per cent, respectively, for $10^8 \Ms$ haloes.  The large scatter in
the MC haloes is created by a wide spread in the gas fraction $f_{\rm
  gas}$ and star formation efficiency $f_\star$ (see Figure
\ref{fig:stats}), where radiation in a gas-poor halo propagates
through a smaller neutral column, resulting in a higher escape
fraction, and vice-versa.  Only five per cent of the ionizing
radiation escape from the largest haloes ($\mvir \ge 10^{8.25} \Ms$)
in the simulation.

\begin{figure}
  \centering
  \includegraphics[width=\columnwidth]{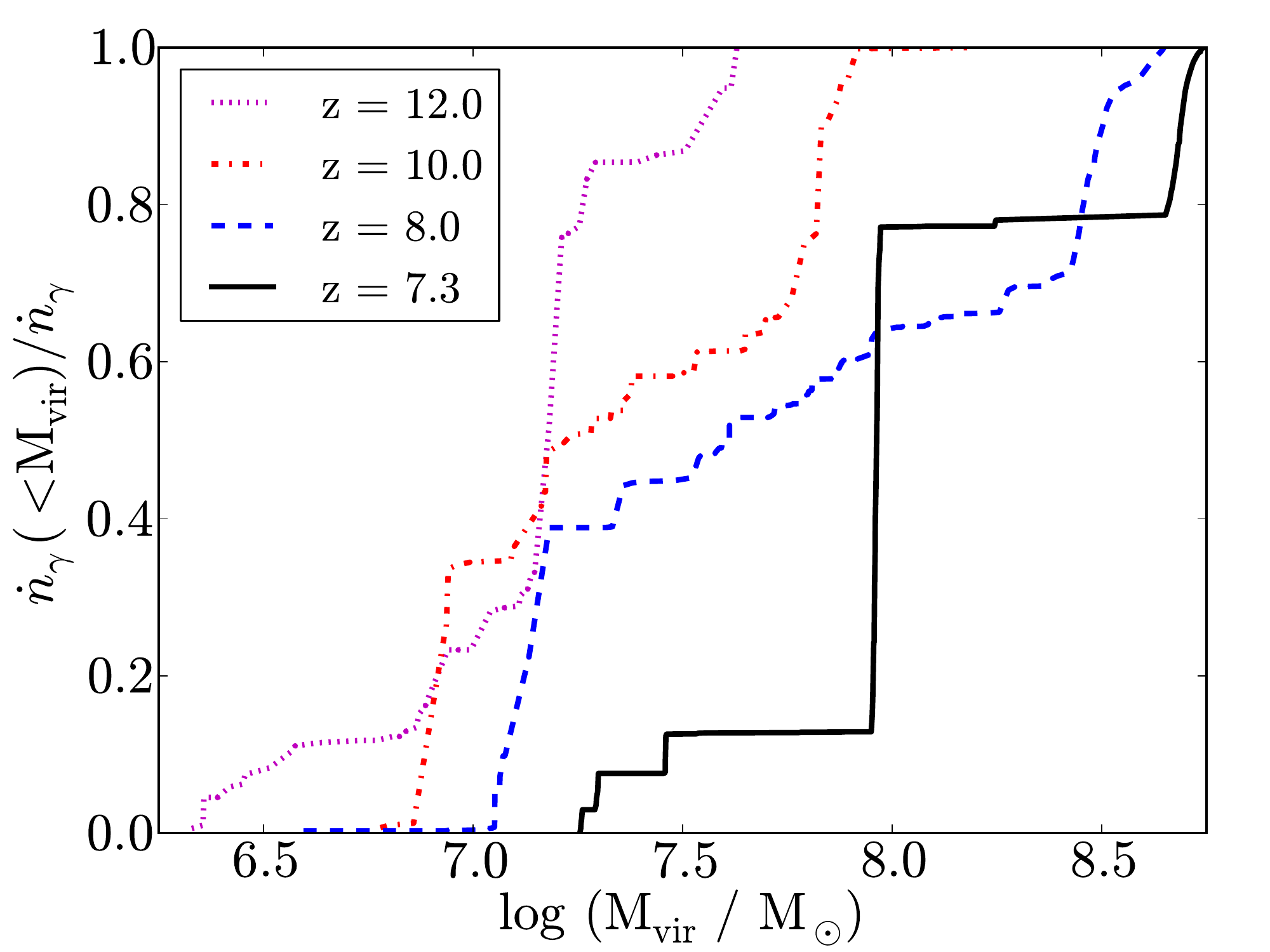}
  \caption{Normalized cumulative probability function of the ionizing
    emissivity of photons that escape into the IGM as a function of
    halo mass at redshifts 12 (dotted), 10 (dot-dashed), 8 (dashed),
    and 7.3 (solid).  The shift to higher halo masses demonstrates
    that low-mass haloes provide a significant fraction of the
    ionising photon emissivity at early times.  The total ionizing
    photon emissivity is $(0.5, 0.9, 3.1, 7.3) \times 10^{50}
    \unit{s}^{-1} \unit{cMpc}^{-3}$ at $z = (12, 10, 8, 7.3)$,
    respectively.}
  \label{fig:fesc-mvir}
\end{figure}

The middle and bottom panels of Figure \ref{fig:fesc} show the
behaviour of the products $\fesc f_\star$ and $\fesc f_\star f_{\rm
  gas}$.  These products are essential ingredients in semi-analytic,
and semi-numerical reionization calculations and numerical simulations
that do not resolve the multi-phase ISM of such small galaxies.  In
Section \ref{sec:stats}, we showed that even the MC haloes can form
stars at an appreciable rate with average values of $f_\star$ just
under 0.01 for typical star-forming minihaloes.  The decrease in
\fesc~dominates the downward trend in these products with respect to
halo mass.  The product $\log \fesc f_\star$ decreases from $-2.8$ to
$-3.6$ ($-2.3$ to $-3.0$) in the halo mass range $10^{6.75} -
10^{8.75} \Ms$ for the mean (luminosity-averaged) values.  Similarly,
the product $\log \fesc f_\star f_{\rm gas}$ decreases from $-3.9$ to
$-4.5$ ($-3.2$ to $-3.9$) for the mean (luminosity-averaged) values in
the same mass range.  This suggests that MC haloes may contribute
significantly to the ionizing photon budget.  To further demonstrate
their contribution, Figure \ref{fig:fesc-mvir} shows the normalized
cumulative escaping ionizing photon emissivity as a function of halo
mass at $z = 12, 10, 8, 7.3$ when the total ionizing emissivity of
escaping photons is $(0.5, 0.9, 3.1, 7.3) \times 10^{50}
\unit{s}^{-1}$ per comoving Mpc$^3$, respectively.  At $z \ge 10$ in
the simulation, low-mass haloes produce the majority of the ionizing
emissivity in the IGM before they cease to form stars because of
either Lyman-Werner or Jeans suppression.  At later times, the $M \ga
10^8 \Ms$ haloes in the simulation start to dominate the ionizing
emissivity, eventually providing nearly 80 per cent at the end of the
simulation at $z=7.3$.  The large jump at $M \sim 10^8 \Ms$ is caused
a single halo undergoing a strong star formation event with a
$\textrm{sSFR} \simeq 30$ and $\fesc \simeq 0.6$ that produces 70 per
cent of the ionising photon budget in the simulation with the
remaining 20 and 10 per cent originating from more massive and less
massive haloes.

\begin{figure}
  \centering
  \includegraphics[width=\columnwidth]{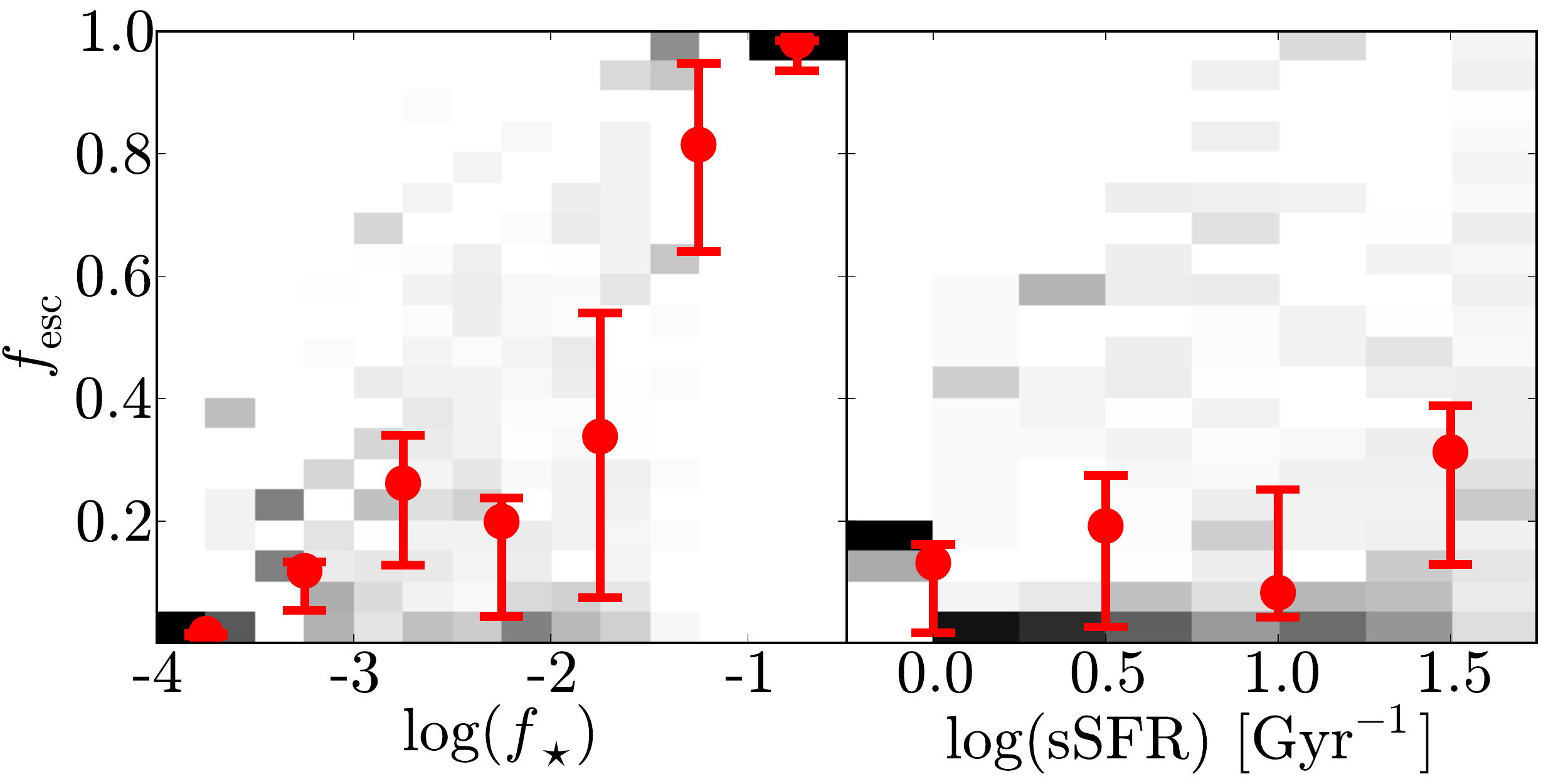}
  \caption{Same as Figure \ref{fig:fesc} but showing the UV escape
    fraction as a function of star formation efficiency $f_\star$ and
    specific SFR.}
  \label{fig:fesc2}
\end{figure}

In principle, the luminosity-weighted average is always higher than
the mean value.  This occurs because, at a fixed halo mass, galaxies
with higher luminosities will ionise their ISM and create ionised
channels out to the IGM more efficiently, boosting \fesc~in such
galaxies.  To explore any correlation between the escape fraction and
star formation, we plot the PDF of \fesc~as a function of the star
formation efficiency $f_\star$ and the specific SFR (sSFR = SFR /
$M_\star$) in Figure \ref{fig:fesc2}.  The mean \fesc~as a function of
these quantities are also listed in Table \ref{tab:fesc2}.  The escape
fraction has a gradual upward trend in the range $f_\star = 10^{-4} -
10^{-2}$, increasing from two per cent in the most inefficient star
forming galaxies to $\sim 20$ per cent.  In galaxies with higher
values of $f_\star$, the escape fraction rapidly increases to nearly
unity in galaxies with $f_\star \ge 0.1$.  This makes physical sense
because if a large fraction of gas forms stars, then there will be
more ionizing photons per baryon in the halo, accelerating the
ionisation of the ISM and thus the escape of any ionizing radiation.
Inspecting Figure \ref{fig:stats}, only the lowest-mass galaxies
exhibit such high efficiencies.  The scatter in this relation is
caused by the variations in the halo's star formation history and
environment.  The behaviour of \fesc~with respect to sSFR shows no
apparent trend, and most of the galaxies \fesc~values lie between
2--40 per cent.

\begin{table}
  \caption{UV escape fractions as a function of star formation
    efficiency and specific star formation rate}
  \label{tab:fesc2}
  \centering
  \begin{tabular*}{0.5\columnwidth}{@{\extracolsep{\fill}}c c}
    \hline\hline
    $\log f_\star$ & $f_{\rm esc}$\\
    \hline\\[-1.5ex]
    --3.75 & $0.02^{+0.00}_{-0.00}$\\[1ex]
    --3.25 & $0.12^{+0.01}_{-0.06}$\\[1ex]
    --2.75 & $0.26^{+0.08}_{-0.13}$\\[1ex]
    --2.25 & $0.20^{+0.04}_{-0.15}$\\[1ex]
    --1.75 & $0.34^{+0.20}_{-0.26}$\\[1ex]
    --1.25 & $0.81^{+0.13}_{-0.18}$\\[1ex]
    --0.75 & $0.99^{+0.00}_{-0.00}$\\[1ex]
    \hline\hline
    sSFR [Gyr$^{-1}$] & $f_{\rm esc}$\\
    \hline\\[-1.5ex]
    0.0 & $0.13^{+0.03}_{-0.11}$\\[1ex]
    0.5 & $0.19^{+0.08}_{-0.16}$\\[1ex]
    1.0 & $0.08^{+0.17}_{-0.04}$\\[1ex]
    1.5 & $0.31^{+0.08}_{-0.18}$\\[1ex]
    \hline
  \end{tabular*}
  \parbox[t]{0.5\textwidth}{\textit{Note:} 
    The mean values are shown for galaxies at all times in 0.5 dex bins.
  }
\end{table}



\begin{figure*}
  \centering
  \begin{minipage}{0.48\textwidth}
    \centering
    \includegraphics[width=\textwidth]{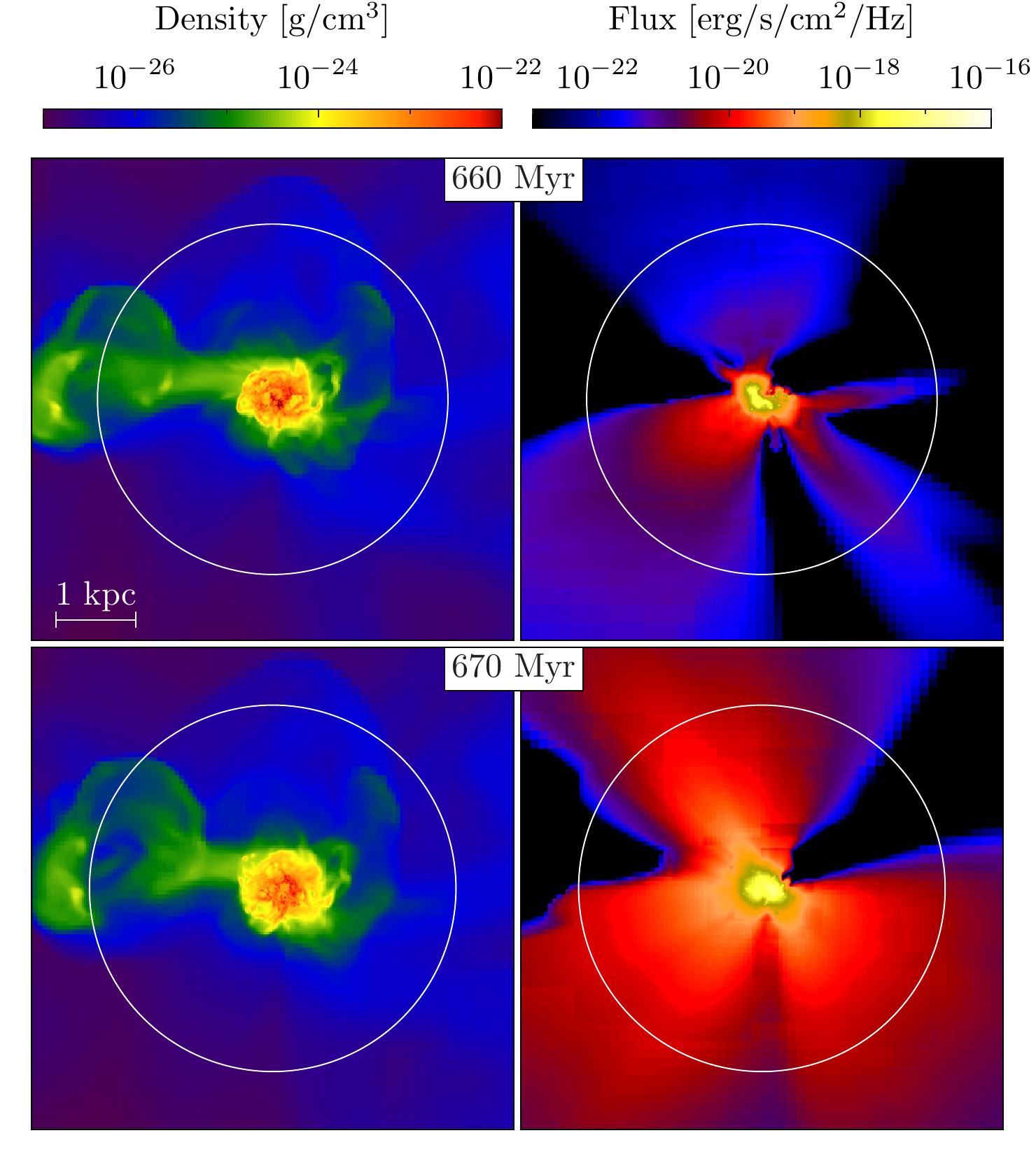}\\
    \vspace{0.4 cm}
    \includegraphics[width=\textwidth]{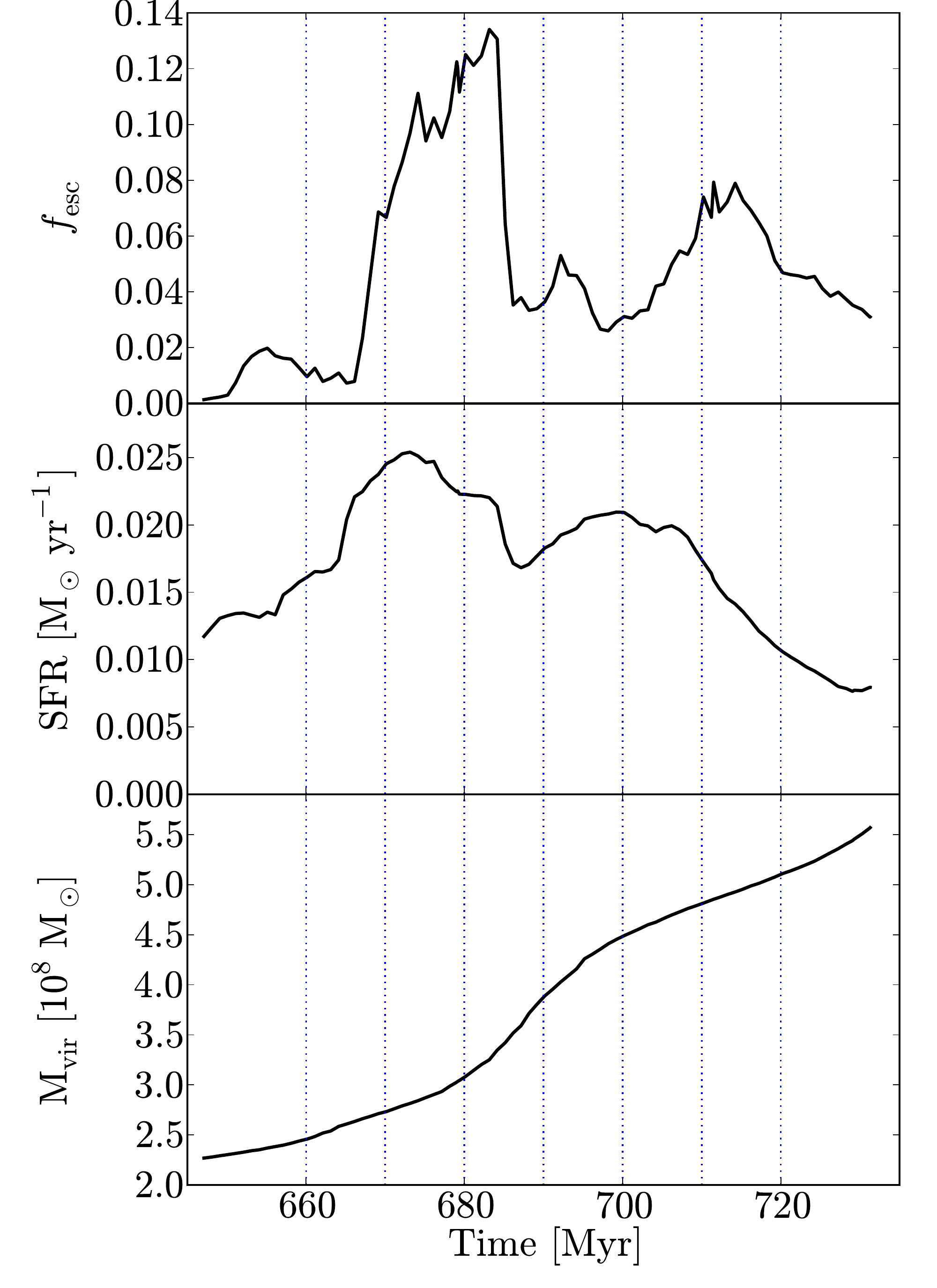}
  \end{minipage}
  \begin{minipage}{0.48\textwidth}
    \centering
    \includegraphics[width=\textwidth]{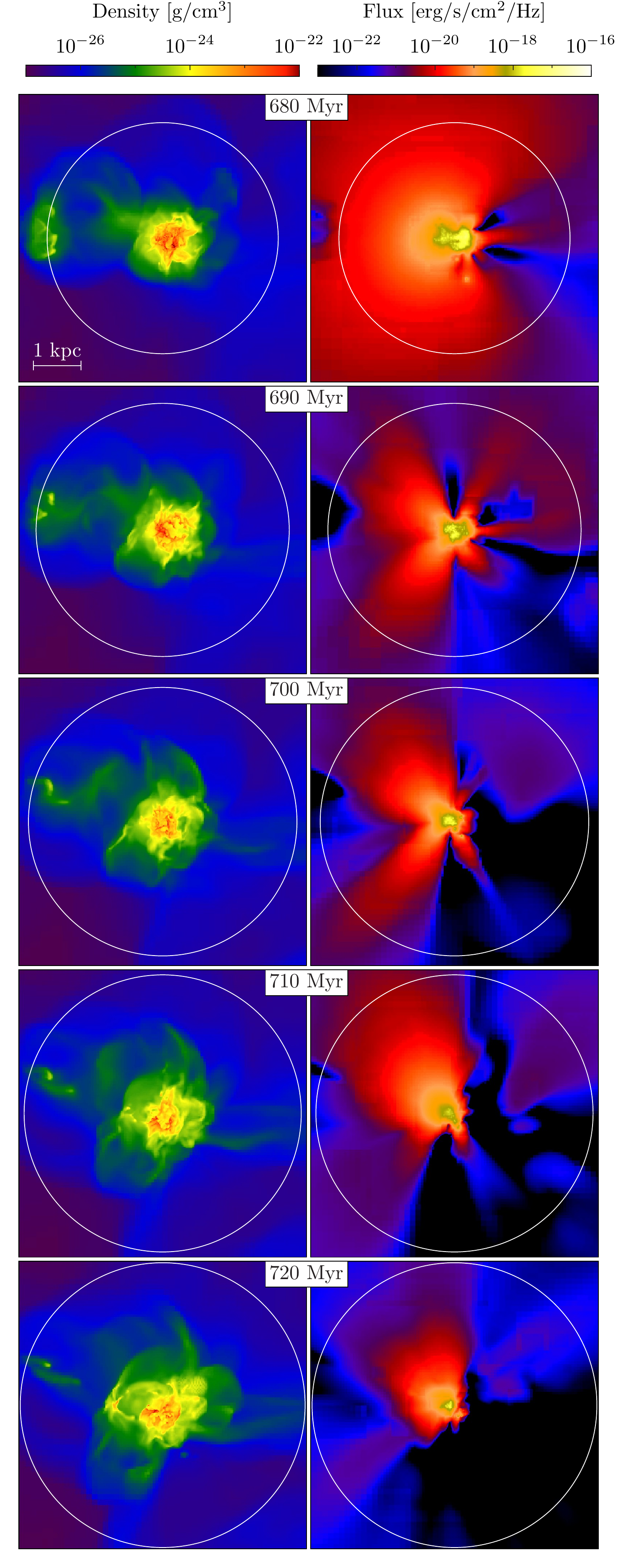}
  \end{minipage}
  \caption{\textit{Lower left:} The evolution of the UV escape
    fraction (top panel), SFR (middle panel), and halo mass (bottom
    panel) of the most massive halo, starting at $z =
    8$. \textit{Remaining panels:} Projections of density-weighted gas
    density (left column) and average UV ($E = 21.6$~eV) flux (right
    column) of the most massive halo every 10 Myr, starting at 660
    Myr, which are marked in the time-evolution plot.  These
    projections shows the variance of the gaseous structure in the
    galaxy and the directions in which UV radiation can escape from
    the halo.  The projections have a field of view of 6 proper kpc
    and a depth of 600 proper pc.  The circles show the virial radius
    at each time.}
  \label{fig:fesc-main}
\end{figure*}

\subsubsection{Time-dependent behaviour and correlation with star
  formation rates}

So far, we have focused on the star-forming properties and UV escape
fractions independent of time.  However, galaxies are very dynamic
with widespread turbulence, \hii~regions growing and then recombining,
and SNe blastwaves propagating through the ISM all occurring while the
halo hierarchically grows.  To explore the time-dependence of the
escape fraction, we focus on the most massive halo that grows from
$2.2 \times 10^8 \Ms$ to $5.5 \times 10^8 \Ms$ during the last 80 Myr
of the simulation.  The lower-left panel in Figure \ref{fig:fesc-main}
shows the escape fraction, SFR, and halo mass as a function of time,
starting at $z=8$.  The remaining panels in Figure \ref{fig:fesc-main}
show projections of density-weighted gas density and average UV flux
at $E = 21.6$~eV inside of this halo at 10 Myr intervals.

This galaxy can sustain star formation at all times, and it
experiences two epochs of stronger star formation.  At $z=8$, the SFR
is $1.1 \times 10^{-2} \,\hsfr$, which then increases to $2.5 \times
10^{-2} \,\hsfr$ 20 Myr later and then decays as the cold dense gas is
disrupted inside the galaxy.  Afterward the gas can cool again and
re-collapses to produce a slightly weaker burst of star formation at
$2.0 \times 10^{-2} \,\hsfr$, and it subsequently steadily decreases
to $0.8 \times 10^{-2} \,\hsfr$ at the end of the simulation.

These two bursts of star formation induce a spike in \fesc, which is
clearly seen in Figure \ref{fig:fesc-main}.  Before the first burst,
the escape fraction varies around 0.01, and then once the burst
occurs, it rapidly increases to 0.13 as the \hii~region partially
breaks out of the halo.  The flux images illustrate the anisotropic
escape of ionizing radiation.  The peak in \fesc~occurs 10~Myr after
the peak in SFR because of the time necessary for the ionization front
to propagate to the virial radius.  The density projections show the
irregular morphology of the ISM and an adjacent filament that provides
smooth cosmological accretion.  This filament is eroded through
photo-evaporation from the radiation originating in the galaxy.

At $t = 680 \,\textrm{Myr}$, a $10^8 \Ms$ halo enters the virial
radius of the most massive halo but enters in a direction that is
nearly perpendicular to the projection.  The additional gas sparks
another burst of star formation at $t = 700 \,\textrm{Myr}$, but also
increases the total \hi~column density between the galaxy and virial
radius.  This additional column and the smaller SFR causes \fesc~to
only reach 8 per cent at its peak, which then decays to 4 per cent as
the SFR decreases.

The image series in Figure \ref{fig:fesc-main} illustrates how the
direction and escape fraction of the radiation can change on
timescales shorter than 10 Myr, arising from the ever-changing
thermal, ionization, and density states of the ISM.  For instance, the
filament to the left of the halo is photo-evaporated from 660 Myr to
680 Myr, and afterwards ionizing radiation can escape in that solid
angle.  These directional and time-dependent properties illustrate how
observations of UV flux blueward of the Lyman break are difficult to
capture even if the UV escape fraction is higher than 10 per cent.
However, they are unimportant for global semi-analytic and
semi-numerical models of reionization when the time-averaged
quantities of \fesc~are adequate.

\subsection{Clumping factor}

\begin{figure}
  \centering
  \includegraphics[width=\columnwidth]{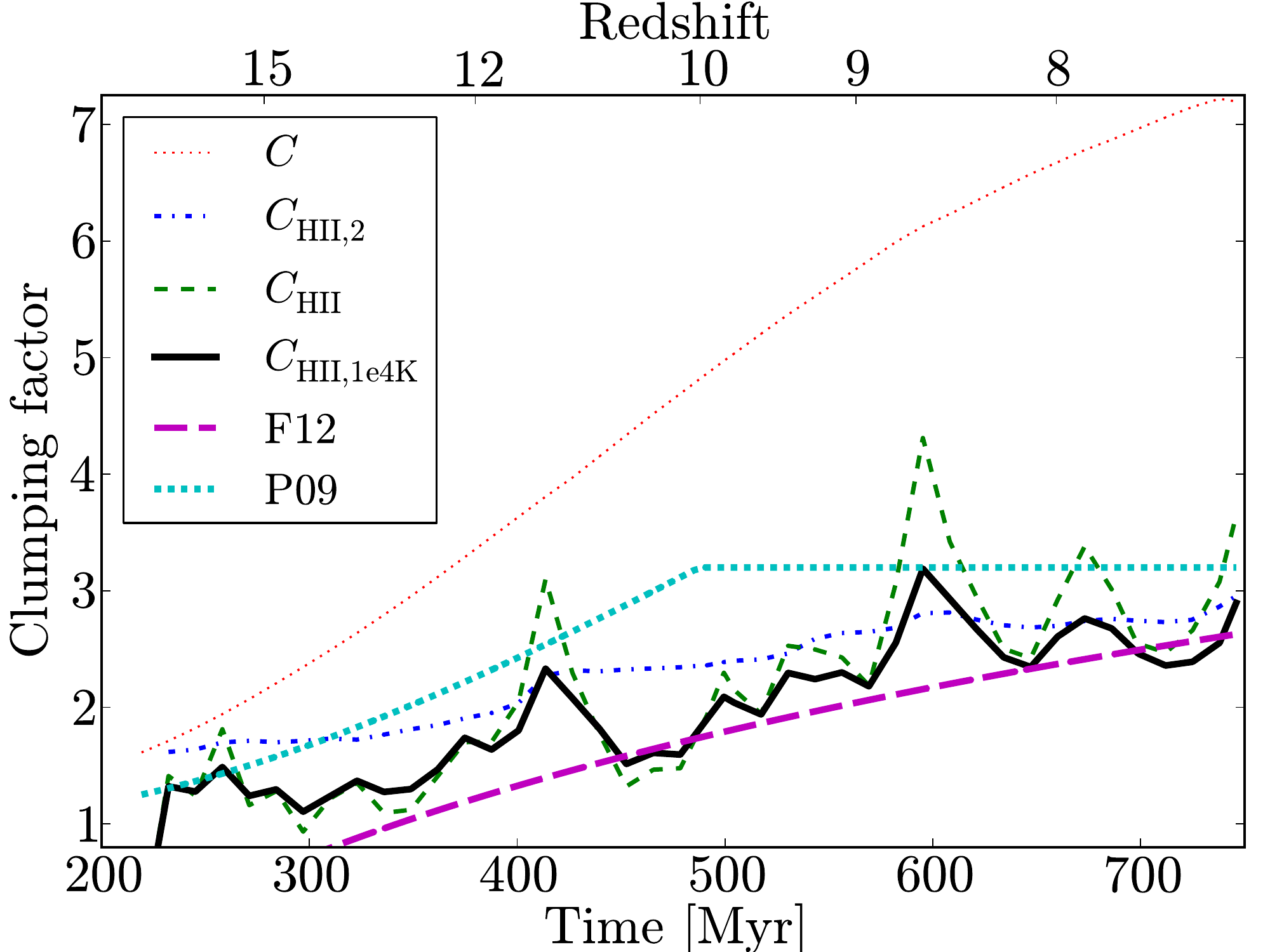}
  \caption{Evolution of the clumping factor as a function of time,
    according to several different definitions, all of which restrict
    the analysis to the diffuse ($\rho/\rho_c < 20$) IGM.  $C$ shows
    the unfiltered clumping factor; $C_{\rm HII,2}$ restricts the
    analysis to ionised ($x > 0.1$) regions; $C_{\rm HII}$ (Equation
    \ref{eqn:clumpHII}) is weighted by $x$; $C_{{\rm HII}, 10^4
      \rm{K}}$ (Equation \ref{eqn:clumpRRmod}) is an observational
    temperature-corrected clumping factor of ionised gas.  The peaks
    in the latter two definitions are caused by inhomogeneous IGM
    being incorporated into an \hii~region and subsequently
    photo-evaporated.  The purple long-dashed and cyan dotted lines
    are fits from \citet{Finlator12} and \citet{Pawlik09},
    respectively.}
  \label{fig:clumping}
\end{figure}

Once ionizing radiation escapes into the IGM, it can create a
cosmological \hii~region.  To remain ionised, the ionization rate must
be greater than the recombination rate.  Recombination in a clumpy IGM
is accelerated in overdense regions because its rate is proportional
to the product of the electron and proton number densities, $n_e n_p$.
This enhancement factor is generally characterised by the clumping
factor $C$ that is traditionally defined as $C \equiv \langle \rho^2
\rangle / \langle \rho \rangle^2$, where the brackets indicate a
volume average.  \citet[][hereafter F12]{Finlator12} explored the
variance of $C$ with its definition, whether it is calculated as
\begin{enumerate}
\item the clumping of all baryons above some threshold density to
  delineate between collapsed objects and the IGM,
\item the clumping of electrons and protons, 
  \begin{equation}
    \label{eqn:clumpHII}
    C_{\rm H II} \equiv \langle n_e n_{\rm H II} \rangle / \langle n_e
    \rangle \langle n_{\rm H II} \rangle,
  \end{equation}
  that can be computed in only ionised regions or be weighted by the
  ionised volume fraction $x$ (see Section 3 in F12 for a discussion),
\item a clumping factor that improves method (ii) by using the
  recombination rate (RR) $\alpha_{\rm B}(T)$ at the temperature of
  the simulated gas \citep{Shull12} and is also weighted by $x$,
  \begin{equation}
    \label{eqn:clumpRR}
    C_{{\rm H II, RR}} \equiv \frac{\langle n_e n_{\rm HII}
      \alpha_{\rm B}(T)\rangle}{\langle n_e \rangle \langle n_{\rm HII} \rangle
      \langle \alpha_{\rm B}(T) \rangle}
  \end{equation}
\item an ``observational temperature-corrected'' clumping factor of
  ionised gas that modifies method (iii) by replacing the
  recombination rate in the denominator with the value at $T = 10^4$~K
  because the observed mean IGM temperature is poorly constrained in
  ionised regions,
  \begin{equation}
    \label{eqn:clumpRRmod}
    C_{{\rm H II}, 10^4 {\rm K}} \equiv \frac{\langle n_e n_{\rm HII}
      \alpha_{\rm B}(T)\rangle}{\langle n_e \rangle \langle n_{\rm HII} \rangle
      \alpha_{\rm B}(10^4 {\rm K})}
  \end{equation}
\end{enumerate}
F12 find that the clumping factor is sensitive to its definition,
varying in the range 2--4 at $z = 6$.  We find similar variations, and
we show them in Figure \ref{fig:clumping}, which depicts the evolution
of $C$, two definitions of $C_{\rm HII}$, $C_{{\rm HII}, 10^4 {\rm
    K}}$, the fit from F12,
\begin{equation}
  \label{eqn:F12}
  x_{\rm HII,v} C_{{\rm HII}, 10^4 {\rm K}} = 9.25 - 7.1 \log_{10} (1+z)
\end{equation}
and the $C_{100}$ fit ($z \sim 10.5$ reionization case) from
\citet{Pawlik09},
\begin{equation}
  \label{eqn:P09}
  C_{\rm HII} = \left\{ \begin{array}{r@{\quad}l} 
      1 + \exp(-0.28z + 3.59) & (z \ge 10)\\
      3.2 & (z < 10)
    \end{array} \right. .
\end{equation}
For all of these definitions of $C$, we restrict our analysis to the
diffuse IGM with $\rho/\rho_c < 20$.

We show two definitions of $C_{\rm HII}$, where $C_{\rm HII}$ is
weighted by the ionised fraction $x$, and $C_{\rm HII, 2}$ is the
clumping factor for partially ionised ($x > 0.1$) gas.  All
definitions of $C$ in the ionised regions generally increase from 1.5
to 3 from redshift 15 to 7.2, agreeing with other works
\citep{Pawlik09, McQuinn11, Shull12, Finlator12, So13}.  The evolution
of $C_{\rm HII,2}$ shows two abrupt increases at $z \sim 11$ and 9
followed by little evolution over the next $\sim 100$ Myr.  Both occur
when a galaxy quickly ionises the surrounding several kpc during a
star formation event that has a high $f_{\rm esc}$, also apparent in
the jumps in ionization fraction $x$ in Figure \ref{fig:ion}.  These
events are more clear in $C_{\rm HII}$ and $C_{{\rm HII}, 10^4 {\rm
    K}}$, where the sudden increases in $C$ occur when the \hii~region
grows and encompasses overdensities that were previously neutral.
After a sound crossing time, these clumps photo-evaporate, and the
clumping factor decreases.  These individual features are caused by
the small simulation volume, where this behaviour in many different
\hii~regions would average out, resulting in a smoothly increasing
clumping factor.  Our clumping factor is consistently higher than the
F12 fit by $\sim 0.25$ on average because we resolve smaller systems
above the cosmological Jeans mass that contribute to IGM clumpiness
\citep{Emberson13}.

\section{Reionisation model}

Small-scale simulations have the benefit of higher resolution, but
they cannot capture the formation of rare large-scale density peaks
and the evolution of global reionization from a statistically complete
sample of galaxies.  Low-luminosity galaxies hosted in MC haloes
should contribute a significant fraction of ionizing photons at very
high redshifts because of their large number densities.  On the other
hand, they are susceptible to negative radiative feedback because
these haloes can be photo-evaporated from external UV radiation and
their gas can be easily expelled by SN explosions.  To assess the
global impact of these smallest galaxies during reionization, we
utilise the mean galaxy properties and escape fractions from our
simulation in a semi-analytical model of reionization.

\subsection{Method}


We compute the evolution of the hydrogen ionised fraction $x$ by
solving
\begin{equation}
  \label{eqn:reion}
  \dot{x} = \frac{\dot{n}_\gamma}{\bar{n}_{\rm H}} -
  \frac{x}{\bar{t}_{\rm rec}},
\end{equation}
where $\dot{n}_\gamma$ is the comoving ionising photon emissivity,
$\bar{n}_{\rm H}$ is the mean comoving hydrogen number density, and
$\bar{t}_{\rm rec} = [C(z) \alpha_B n_{\rm H} (1 + Y/4X)
(1+z)^3]^{-1}$ is an effective recombination time for a fully ionized
hydrogen gas \citep[see][for a discussion]{So13}.  $X = 0.76$ and $Y =
1-X$ are the hydrogen and helium number fractions, respectively
\citep{Shapiro87_HII, Madau99}.  We use the clumping factor $C_{\rm
  HII}$ from \citet[Equation \ref{eqn:P09};][]{Pawlik09} and start
with an initial ionised fraction of $10^{-5}$ at $z=30$.  We integrate
this equation until the volume is completely ionised, giving a full
reionisation history from which we can calculate the optical depth due
to Thomson scattering,
\begin{equation}
  \tau_e = \int_0^\infty dz \frac{c(1+z)^2}{H(z)} x(z) \sigma_T
  \bar{n}_H (1 + \eta Y/4X),
\end{equation}
where $H(z)$ is the Hubble parameter, $\sigma_T$ is the Thomson
cross-section.  We assume that the helium is singly ionised ($\eta =
1$) at $z > 3$ in the same volume fraction as hydrogen and doubly
ionised ($\eta = 2$) at later times.  Comparing the calculated
$\tau_e$ of our models to the observed values from WMAP and {\it
  Planck}, $\tau_e = 0.089^{+0.012}_{-0.014}$, can place constraints
on the timing of reionisation and the ionising emissivity.

\subsubsection{Ionizing emissivity}

Most of the uncertainties are hidden in the ionising photon emissivity
(Equation \ref{eqn:ngamma_halo}).  For a halo mass independent model,
$\dot{n}_\gamma = f_{\rm esc} f_\gamma \dot{\rho}_\star$, where
$\dot{\rho}_\star$ is the SFR density, and $f_\gamma$ is the photon to
stellar baryon number ratio.  However, we have shown that the escape
fraction and SFRs are strong functions of halo mass $M$; therefore,
we consider
\begin{equation}
  \label{eqn:ngamma}
  \dot{n}_\gamma = \int_{M_{\rm min}}^\infty f_{\rm esc} f_{\gamma}
  (f_{\rm gas} f_{\star} \dot{f}_c \mvir) \, d\mvir,
\end{equation}
where {\it all} of the factors inside the integral are functions of
halo mass, and the product inside the parentheses is the cosmic SFR
density in halo masses between $M$ and $M+dM$.  We take $M_{\rm min} =
10^{6.25} \Ms$ from our simulation.  Here $\dot{f}_c$ is the
time-derivative of the collapsed fraction that is calculated with the
ellipsoidal variant of Press-Schetcher formalism \citep{PS74, Sheth01}
with the same cosmological parameters used in the simulation.  

We compute $\dot{n}_\gamma$ by discretising the integral in Equation
(\ref{eqn:ngamma}) into mass bins of $\log \Delta (\mvir/\Ms) = 0.5$
in $\log (\mvir/\Ms) \in [6.25, 12.75]$.  At $\log(\mvir/\Ms) < 8.75$,
we utilise the luminosity weighted mean values of $f_{\rm gas} f_\star
f_{\rm esc}$ from the simulation (Table \ref{tab:fesc}).  For more
massive haloes that are not sampled by our simulation, we assume a
mass-independent values \citep[cf.][]{Alvarez12}, $(f_\star, f_{\rm
  esc}, f_{\rm gas}) = (0.03, 0.05, \Omega_b/\Omega_m)$.

Metal-poor stars can produce up to a factor of two more ionising
photons per stellar baryon, compared to their solar metallicity
counterparts, because of higher surface temperatures.  To account for
this effect, we utilise the photon to stellar baryon ratio
$f_\gamma(Z)$ from \citet[their Equation 1]{Schaerer03}.  Then we can
express $f_\gamma = f_\gamma[Z(\mvir)]$ through two relations: (i) the
local dwarf galaxy metallicity-luminosity relation \citep{Woo08},
\begin{equation}
  \label{eqn:ZL}
  \log(Z) = -3.7 + 0.4 \log \left( \frac{M_\star}{10^6 \Ms} \right),
\end{equation}
and (ii) our simulated $M_\star - \mvir$ mean values (Table
\ref{tab:stats}), resulting in an decreasing $\gamma$ with increasing
\mvir.

Lastly, we include emissivity from Population III stars in one model.
They only form in metal-free haloes that are susceptible to LW
feedback and metal enrichment.  Simulations that include these physics
show the SFR increasing for the first $\sim$100~Myr, and afterwards
the SFR becomes approximately constant prior to reionization
\citep[e.g.][]{Wise12_Galaxy, Xu13}.  Therefore, we assume a constant
Pop III SFR of $5 \times 10^{-5}\,\sfr$, starting at $z=30$, with
$f_\gamma =$ 60,000 \citep{Schaerer02}.  This corresponds to
$\dot{n}_\gamma = 1.1 \times 10^{50} \pem$.  Note that this emissivity
does not enter the integral in Equation (\ref{eqn:ngamma}) but adds a
constant term to $\dot{n}_\gamma$.

\subsubsection{Dwarf galaxy suppression}

The suppression of star formation occurs in low-mass haloes that are
susceptible to negative feedback through photo-dissociation, gas
blowout, and/or photo-evaporation.  We showed in Figure
\ref{fig:occupy} that the halo-galaxy occupation fraction $f_{\rm
  host}$ decreases from an initial value $f_0$ to a final value $f_1$
after a sound crossing time \citep[cf.][]{Shapiro04, Sobacchi13}
\begin{equation}
\label{eqn:sc}
t_{\rm sc} \approx 200 \unit{Myr} \left( \frac{M}{10^8 \Ms}
\right)^{1/3} \left( \frac{1+z}{10} \right)^{-1} \left(
  \frac{\Omega_{\rm m} h^2}{0.15} \right)^{-1/3}.
\end{equation}
We apply this suppression when the filtering mass $M_{\rm F}(t_{\rm
  F}) > \mvir$ by multiplying $\dot{n}_\gamma$ by
\begin{equation}
  \label{eqn:smooth}
  f_{\rm host}(M,t) = A - B\,
  \textrm{tanh} \left( \frac{t-t_{\rm F}-t_{\rm sc}/2}{t_{\rm sc}/3}
  \right),
\end{equation}
where $A = (f_0+f_1)/2$ and $B = (f_0-f_1)/2$, which provides a
functional fit to the simulated values of $f_{\rm host}$ that are
shown in Figure \ref{fig:occupy}.  To calculate the filtering mass
(Equation \ref{eqn:filter}), we use a simple model that uses the Jeans
mass $M_{\rm J}$ of an ionised IGM with $T = 10^4 \unit{K}$ and $\rho
= \Omega_b \rho_c$.

Figure \ref{fig:occupy} shows that star formation is suppressed in
galaxies that cannot cool through atomic transitions.  Thus, we only
use this transition for the mass bins centered on $\log(\mvir/\Ms) =
(7.0, 7.5, 8.0)$, where we use $f_0 = (0.3, 0.5, 1.0)$ and $f_1 =
(0.01, 0.15, 0.5)$, respectively.  For $\log(\mvir/\Ms) = 6.5$, we use
$f_{\rm host} = 2.4 \times 10^{-4}$, the time-averaged value, because
these extremely low-mass galaxies are sporadic and rare (see
\S\ref{sec:stats}).  Lastly, we do not alter the emissivities of the
atomic cooling haloes with $\log(\mvir/\Ms) > 8.25$.

\subsection{Results}

%
%

\begin{figure}
  \centering
  \includegraphics[width=\columnwidth]{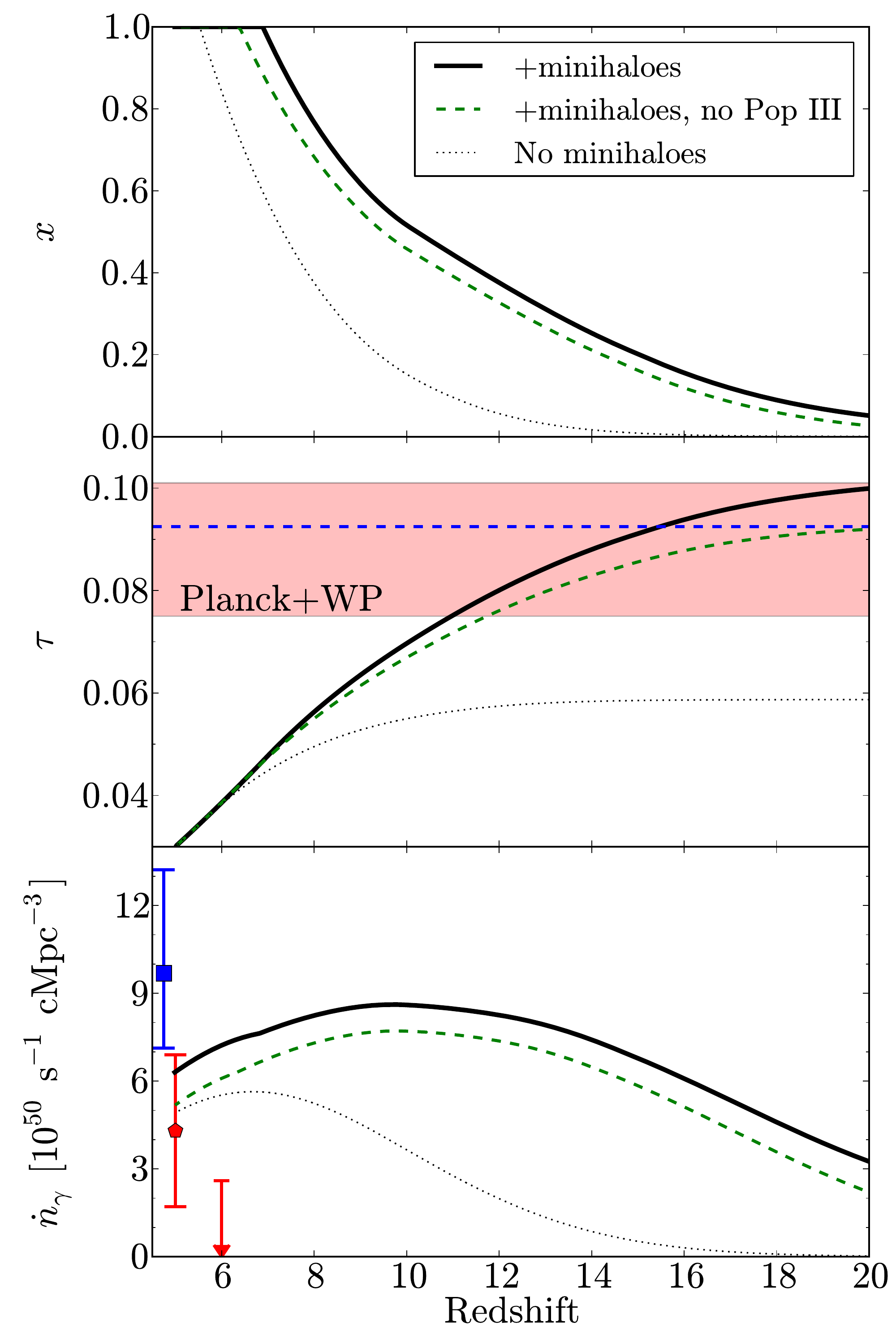}
  \caption{Ionization histories for models that include only atomic
    cooling haloes (dotted lines), adding Pop II star-forming
    minihaloes (dashed lines), and including Pop III star formation
    (solid lines).  The luminosity-weighted mean of the escaping
    ionising luminosity shown in Tables \ref{tab:stats} and
    \ref{tab:fesc} (solid lines), are used.  \textit{Top panel:}
    Evolution of ionised mass fraction.  \textit{Middle panel:}
    Corresponding Thomson scattering optical depth with the
    Planck+WMAP9 best fit and 1-$\sigma$ errors shown.  \textit{Bottom
      panel:} Comoving ionising photon emissivity that escape into the
    IGM.  The points denote constraints from the transmissivity
    through the \lya~forest from \citet[red pentagon and upper
    limit;][]{Kuhlen12} and \citet[blue square;][]{Becker13}.  It
    should be noted that in the range $z = 2-5$, \citeauthor{Becker13}
    found ionising photon emissivities a factor of $\sim 2$ higher
    than \citeauthor{Kuhlen12}.}
  \label{fig:reion}
\end{figure}

Here we consider three models that progressively add more ionising
radiation sources: (i) star formation only in haloes with masses
$\mvir > 10^{8.25}\Ms$ that are approximately atomic cooling haloes,
(ii) plus star formation in minihaloes with masses $\mvir >
10^{6.25}\Ms$, and (iii) plus Pop III star formation.  Figure
\ref{fig:reion} shows the main results from our semi-analytic
reionization models.

The model that restricts star formation to atomic cooling haloes
reionises the universe at $z = 5.5$, resulting in $\tau_e = 0.059$,
which is typical of models either only considering atomic cooling
haloes with low escape fractions that are similar to local dwarf
galaxies or only including observable galaxies at $z \ga 6$
\citep[e.g., see the $M_{\rm UV} = -17$ model of][]{Robertson13}.  To
explain the discrepancy between the observed values of $\tau_e$ and
the models that include only currently observable galaxies, high
escape fractions may be invoked \citep[e.g.][]{Kuhlen12, Alvarez12,
  Robertson13} and other sources of ionising photons, such as X-ray
binaries \citep{Power13, Fragos13} and Pop III stars \citep{Ahn12}.
\citeauthor{Kuhlen12} and \citeauthor{Robertson13} both used a
mass-independent description of the star formation efficiency and
escape fraction, but \citeauthor{Alvarez12} presented a simple model
that delineated between low- and high-mass dwarf galaxies with $f_{\rm
  esc} = (0.8, 0.05)$, respectively.  None of these recent studies
considered minihaloes because they are very susceptible to negative
feedback; however, our simulation shows that they can indeed form a
non-negligible amount of stars with high escape fractions.

\begin{figure}
  \centering
  \includegraphics[width=\columnwidth]{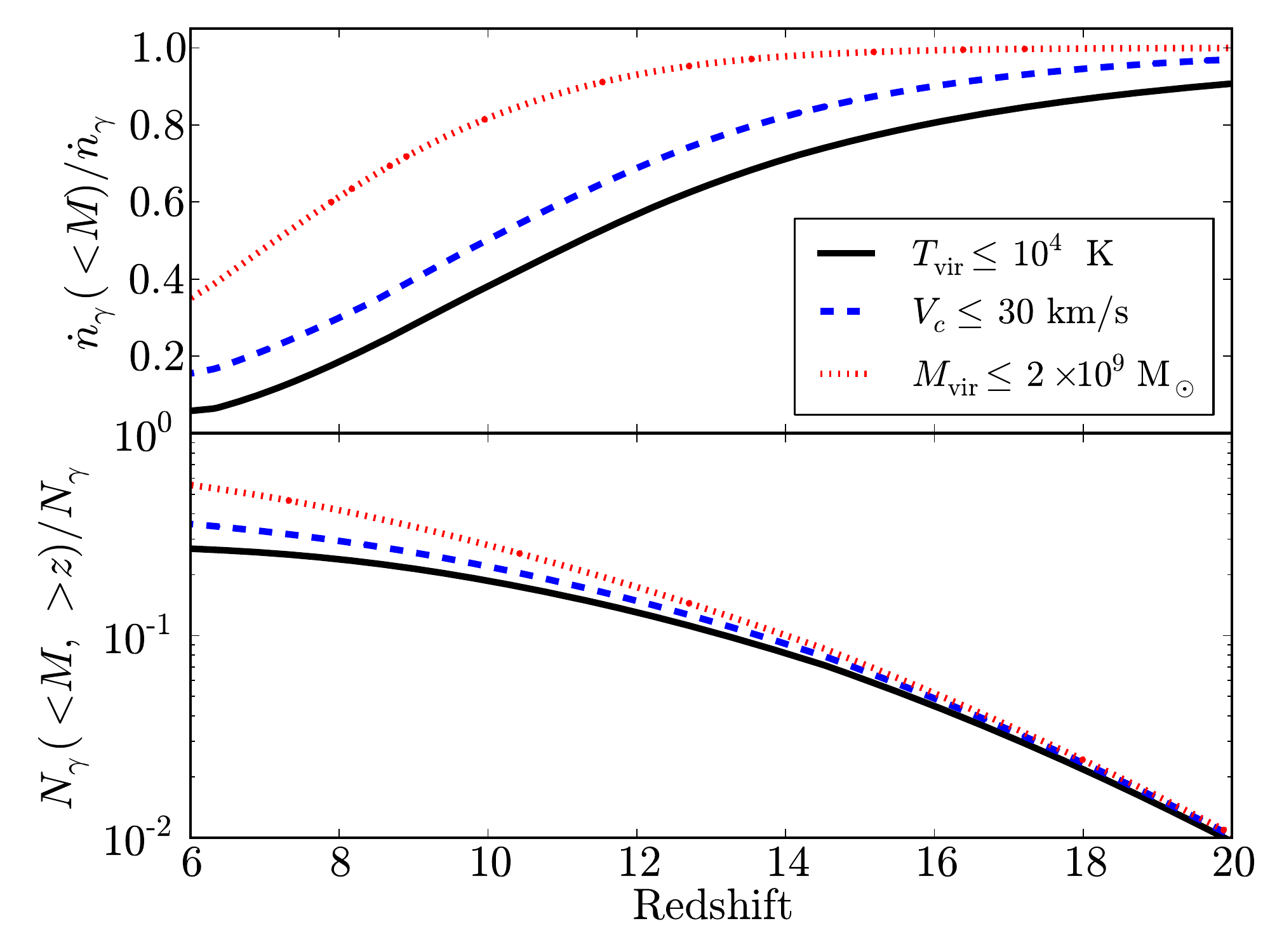}
  \caption{{\it Top panel:} The fractional instantaneous ionising
    emissivity from haloes below three mass thresholds typically used
    for the suppression of star formation, $T_{\rm vir} = 10^4
    \unit{K}$ (solid black), $V_c = 30 \kms$ (blue dashed), and
    $M_{\rm vir} = 2 \times 10^9 \Ms$ (red dotted).  Low-mass haloes
    dominate the photon emissivity at high redshifts, producing an
    ionisation fraction of 20 per cent by $z = 14$, which are
    photo-suppressed at lower redshifts.  {\it Bottom panel:} The
    cumulative fraction of ionising photons emitted from galaxies
    contained in haloes below the same thresholds as the top panel.
    These data demonstrate that even the metal cooling haloes
    contribute nearly 30 per cent to the total photon budget and
    should not be neglected in reionisation calculations.}
  \label{fig:nion_frac}
\end{figure}

Our model (ii) includes such low-luminosity galaxies, which boosts
$\tau_e$ to 0.093 that is very close to the best-fit value from {\it
  Planck} of $\tau_e = 0.0925$ illustrated in the middle panel of
Figure \ref{fig:reion}.  This model reionises the universe at $z =
6.4$.  The bottom panel of Figure \ref{fig:reion} shows that the
ionising photon emissivity is already $2 \times 10^{50} \pem$ at
$z=20$ compared to the negligible emissivity ($2 \times 10^{48} \pem$)
when minihaloes are not included.  It continues to rise until it
reaches $8 \times 10^{50} \pem$ at $z \sim 10$.  Eventually these
galaxies are photo-suppressed, and the total ionising emissivity
converges to model (i) at $z \sim 6$ that are consistent with
observational constraints \citep{Kuhlen12, Becker13}.  These smallest
galaxies have $\fesc \sim 0.5$, and their host haloes collapse much
earlier than the atomic cooling haloes, leading to the boost at very
high redshifts.  The top panel of Figure \ref{fig:reion} shows this
early time behaviour produces a more extended reionisation history
with an ionised volume fraction of $x = 0.2$ by $z = 14$ and $x = 0.5$
by $z = 9.4$.  The top panel of Figure \ref{fig:nion_frac} shows the
instantaneous fractional emissivity from haloes below three different
mass thresholds that are often used for suppression, $T_{\rm vir} =
10^4 \unit{K}$, $V_c = 30 \kms$, and $M_{\rm vir} = 2 \times 10^9
\Ms$.  The MC haloes provide 75 per cent of the ionising emissivity at
$z = 15$, dropping to 10 per cent by $z = 7$, as they are
photo-suppressed, and larger galaxies are the main source of
reionisation at later times.  Similar trends are seen in the other two
halo mass thresholds, where at $z = 10$, haloes with $V_c \le 30 \kms$
and $M_{\rm vir} \le 2 \times 10^9 \Ms$ produce 50 and 80 per cent of
the ionising emissivity, respectively.  The bottom panel of Figure
\ref{fig:nion_frac} shows the cumulative fraction of ionising
radiation that originates from haloes below the same halo mass
thresholds.  The key point to take from these results at $z = 6$ is
that 27, 36, and 56 per cent of the total ionising photon budget are
created in haloes with $T_{\rm vir} \le 10^4 \unit{K}$, $V_c \le 30
\kms$, and $M_{\rm vir} \le 2 \times 10^9 \Ms$, respectively.  Thus,
we conclude that \textit{the lowest luminosity galaxies play an
  integral role in the reionisation of the universe} but are
eventually suppressed as they are engulfed by an increasing UV
radiation field.

Short-lived Pop III stars provide an additional source of ionising
photons, where they ionise the surrounding few kpc, and then this
\hii~region quickly recombines.  However in our model, they have a
low-level contribution to the ionising photon budget, increasing the
$\dot{n}_{\rm ion}$ by $1.1 \times 10^{50} \pem$.  This increases
$\tau_e$ to 0.106 that is in the upper 1-$\sigma$ bound of the {\it
  Planck} results and results in complete reionisation at $z = 6.9$.

\begin{figure}
  \centering
  \includegraphics[width=\columnwidth]{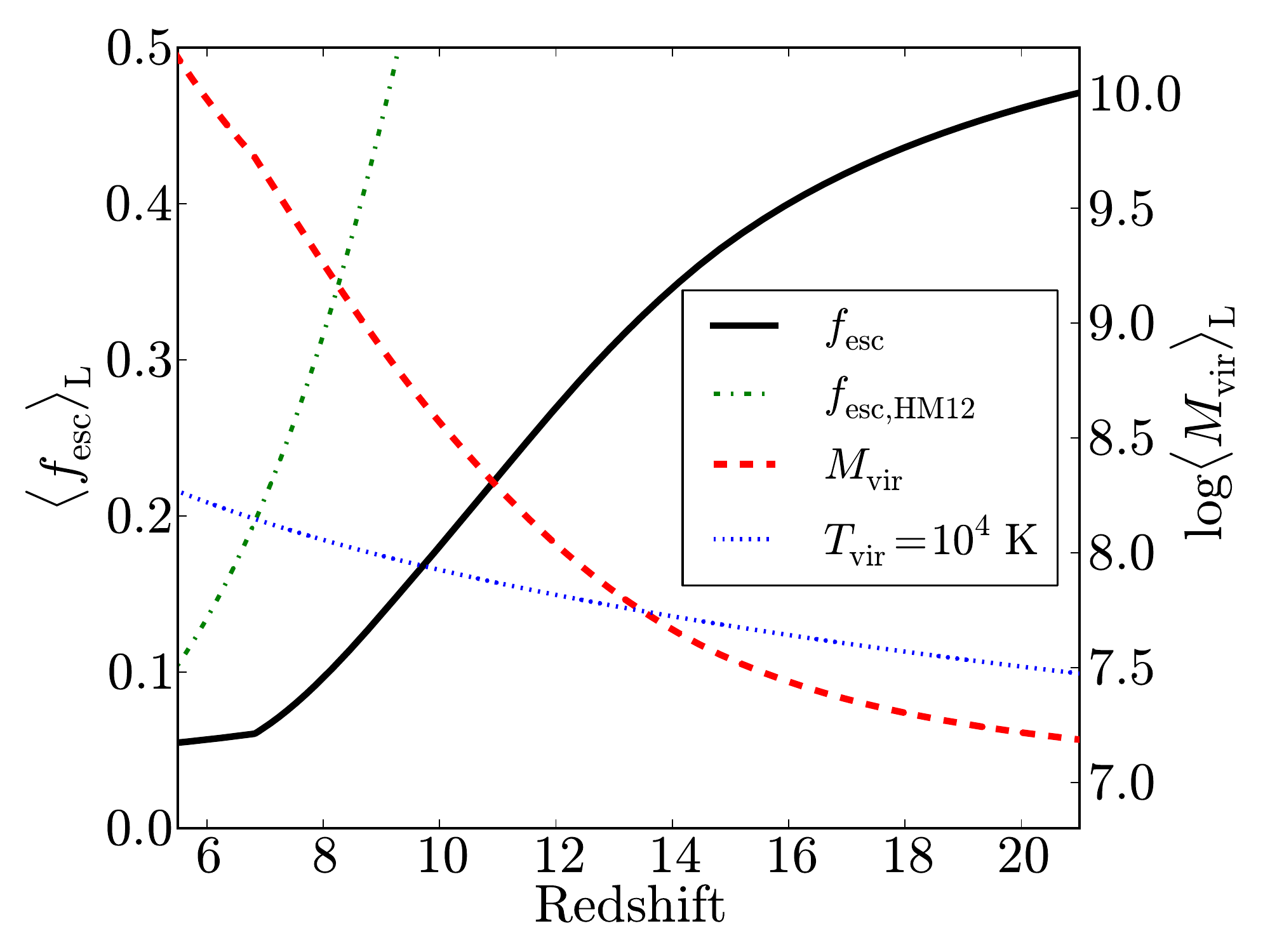}
  \caption{Ionizing luminosity weighted mean values of the ionising
    photon escape fraction (black solid) and halo mass (red dashed) in
    the model that includes minihaloes and Pop III star formation.  For
    comparison, the escape fraction from \citet{Haardt12} (green
    dash-dotted) increases much more rapidly with redshift, and the
    blue dotted line shows the virial mass of a $T_{\rm vir} = 10^4
    \unit{K}$ halo is shown, taking the mean molecular weight $\mu =
    0.6$.}
  \label{fig:fescz}
\end{figure}

For mass-independent models of reionisation, it is useful to consider
luminosity-weighted mean value of \fesc~and the host halo mass, which
are shown as a function of redshift in Figure \ref{fig:fescz}.  At
early times, the low-luminosity galaxies are not suppressed and are
the biggest sources of the ionising photon budget, which is apparent
by the mean halo mass of $10^{7.2} \Ms$ and $\fesc = 0.46$ at redshift
20.  As these galaxies are suppressed, the halo mass scale increases
to $\sim 10^{10} \Ms$ by redshift 6 while the mean escape fraction
gradually drops to $\fesc = 0.05$, which is the assumed escape
fraction of these larger galaxies.  This time-dependent behaviour is
generally true for any model that considers multiple galaxy
populations that are suppressed and unaffected by feedback mechanisms
\citep[cf.][]{Alvarez12}; however, the evolution of \fesc~is much
slower than the one assumed in \citet{Haardt12} as shown in Figure
\ref{fig:fescz}.

Our results are well-fit by analytic functions and can be easily
utilised in reionisation models.  Fitting the time-dependence of
\fesc~with a pure Gaussian had errors up to 20 per cent at early times
and is better fit with a Voigt profile,
\begin{equation}
  \label{eqn:fesc}
  \langle f_{\rm esc} \rangle_{\rm L} (z) = 0.54 - 544 \, G(z,\sigma,z_0) \,
  L(z,\Gamma,z_0),
\end{equation}
where $G$ and $L$ are Gaussian and Lorentzian distribution functions
that are normalised to unity, respectively.  Our best-fit model has
$\sigma = 19.6$, $\Gamma = 7.26$, and $z_0 = 5.8$ and is only valid
for $z > z_0$.  Prior to reionisation, the luminosity-weighted halo
mass scale asymptotes to linear relations at early and late times and
can be fit by a rotated hyperbola.  But for computational convenience,
we can fit the behavior with
\begin{equation}
  \label{eqn:mvir}
  \log \langle M_{\rm vir} \rangle_{\rm L} (z) = 7.0 + (0.25 + 2.96
    \times 10^{-4} z^{3.16})^{-1}.
\end{equation}
Both fits are good to within 3 per cent at $z > 5.8$, where our
$\fesc(z)$ fit reaches a minimum.  These fits encapsulate the effects
of suppression of the lowest mass dwarf galaxies that have high escape
fractions.  Coupled with our mass-dependent dwarf galaxy properties
presented in \S\ref{sec:results}, this resulting reionisation history
demonstrates the role of high-redshift, photo-suppressible dwarf
galaxies during reionisation.

\section{Discussion}
\label{sec:discuss}

We have demonstrated that low-luminosity galaxies play a key role
during the early phases of reionisation.  The exact properties of a
particular high-redshift dwarf galaxy most critically depend on the
star formation history and the ionization and thermal properties of
its local environment.  For instance, the magnitude of radiative and
SN feedback in their progenitors will regulate the gas and
metal content of the galaxy, thus affecting the strength of radiative
cooling and star formation in the dwarf galaxy.

As previous studies have shown \citep[e.g.][]{Clarke02, Wise09,
  Fernandez11, Benson13}, ionising radiation mainly escapes through
low-density channels that are created by ionisation fronts or SNe.
The location and strength of star formation and galaxy morphology can
both influence the UV escape fraction.  Thus, resolving star-forming
clouds and the multi-phase ISM are necessary requirements for any
simulation that aims to measure the escape fraction.  Furthermore,
environmental properties, in particular, the incident radiation can
affect the formation of the first galaxies and their escape fractions.
First, any Lyman-Werner radiation will suppress \hh~formation and thus
Pop III stars in the progenitors.  If Pop III star formation is
delayed until haloes have $M \ga 10^7 \Ms$, the \hii~region and SN
blastwave may fail to breakout of the halo, possibly leading to prompt
metal-enriched star formation \citep{Whalen08_SN, Ritter12}.  Second,
any external ionising radiation will photo-evaporate the gas in
low-mass haloes over a sound-crossing time.  However,
photo-evaporation may boost the escape fraction for a short period if
the halo is already hosting active star forming regions.  If the outer
layers of gas are photo-evaporated by an external source, the neutral
hydrogen column density between the halo center and the IGM will
decrease, causing an increase in escape fraction in principle.  On the
other hand, this reduces the amount of gas available for future star
formation in small galaxies.  If the halo does not accrete additional
gas from smooth accretion or a merger within a sound-crossing time,
the SFR and escape fraction will steadily decrease, where we showed a
similar case of a correlation between SFR and escape fraction in
Figure \ref{fig:fesc-main}.

Although the SFRs in the lowest-mass galaxies are small, their roles
in reionisation are not insignificant.  Now we turn our attention to
the implications of our results on reionisation, their observational
prospects, differences with previous work, and caveats of our study.

\subsection{Implications for reionisation}


The idea of the faintest galaxies providing the majority of the
ionising photon budget of reionisation is not a new one.  However, our
work provides convincing evidence that stars form in these galaxies
before becoming photo-suppressed by internal stellar feedback and
external radiative feedback.  Their star formation efficiencies are
not significantly lower than atomic cooling haloes, but their low
masses lead to low SFRs of $10^{-3} - 10^{-4}~\hsfr$ with about half
of the ionising radiation escaping into the IGM.  Combined with a high
number density of their host haloes, their contribution to
reionisation should not be neglected, especially at very high redshift
($z \ga 10$).  Some of these galaxies are initially embedded in a cool
and neutral ISM, most likely in a relic \hii~region from Pop III
stars, but as global reionisation proceeds, they are likely embedded
in larger \hii~regions, slowly being photo-evaporated over a sound
crossing time.  Then larger galaxies that are hosted by atomic cooling
haloes provide the majority of ionising radiation, as shown in Figure
\ref{fig:fescz}.  In a statistical sense, there will probably be no
detectable transition in the reionisation or SFR history between these
two source types because they form coeval.  A significant fraction of
galaxies that are initially photo-suppressed will likely host star
formation shortly afterward because rapid mass accretion rates at
high-redshift will allow for efficient cooling and star formation even
in the presence of a UV radiation field.

This scenario leads to an extended period of reionisation, where the
universe has an ionised mass fraction of 10 and 50 per cent by $z \sim
17$ and $z \sim 10$, respectively, eventually becoming completely
reionised by $z \sim 6.5$ (see Figure \ref{fig:reion}).  Such an
extended ionisation history produces a Thomson scattering optical
depth $\tau_e = 0.093$ and an ionising emissivity that are consistent
with {\it Planck} and \lya~forest observations, respectively.  We
stress that the faintest galaxies should not be overlooked in
reionisation models and provide the key to matching the latest
ionisation constraints, resolving any tension between CMB observations
and \lya~forest observations at $z \sim 6$.

\subsection{Observational prospects}

\begin{figure}
  \centering
  \includegraphics[width=\columnwidth]{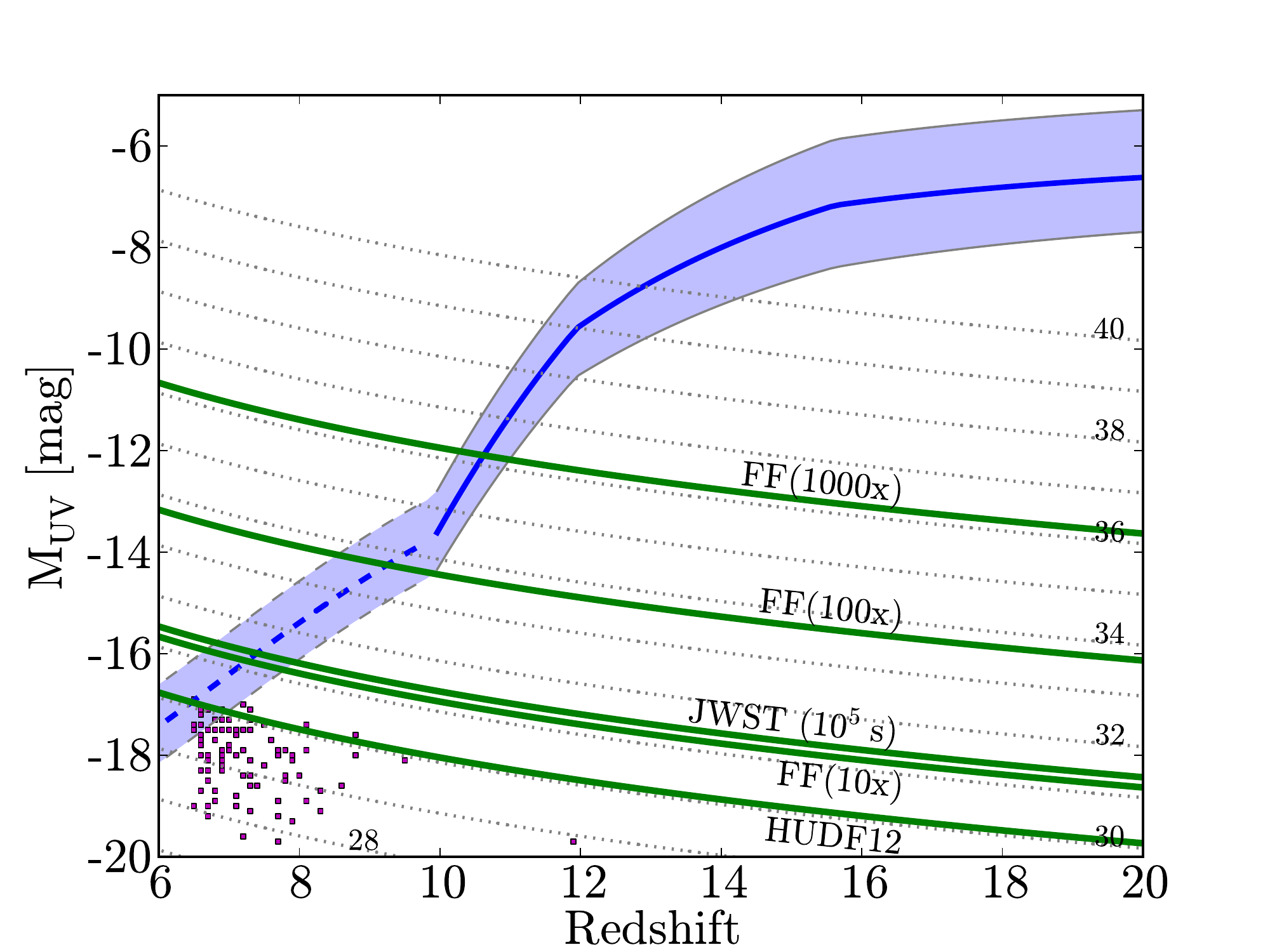}
  \caption{Absolute UV magnitude at 1500~\AA~for the typical dwarf
    galaxy contributing to reionisation.  The solid line uses the mean
    $M_\star$--$M_{\rm vir}$ relation shown in Table \ref{tab:stats}
    to convert the luminosity-weighted average halo mass in Figure
    \ref{fig:fescz}, where the shaded areas show the 1-$\sigma$
    deviations.  The dashed lines extrapolate to halo masses $M \ge
    10^{8.5}\Ms$.  The thin dotted lines show the apparent magnitude
    that are denoted above the lines, where we have assumed a constant
    K-correction of --2.  The solid green lines delineate the
    detection limits for the HUDF12 ($M_{\rm lim} = 30.1$), Frontier
    Fields ($M_{\rm lim} = 28.7 + 2.5\log\mu$), and JWST ultra-deep
    ($M_{\rm lim} = 31.4$) campaigns, where $\mu$ is the magnification
    factor.  Lastly, the magneta squares show the galaxies observed in
    the HUDF12 campaign from the \citet{McLure13} $z \ge 6.5$ robust
    sample.}
  \label{fig:obs}
\end{figure}

Figure \ref{fig:obs} shows the absolute magnitude at 1500~\AA~of a
typical dwarf galaxy that contributes to reionisation as a function of
redshift and compare them to the detection limits of the HUDF12,
Frontier Fields, and JWST ultra-deep campaigns.  Here we take the
luminosity-averaged halo mass, shown in Figure \ref{fig:fescz} and
show the average and 1-$\sigma$ deviations of $M_{\rm UV}$ from Table
\ref{tab:stats}.  Even at $z < 10$, the continuum radiation from these
typical galaxies is below the detection limits in the HUDF12 campaign.
However with sufficient magnification from gravitational lensing,
observations from the \textit{Frontier Fields} may detect such a
galaxy population that contributes the majority of radiation to
reionisation \citep[e.g.,][]{Mashian13, Coe14}.

Recall that our calculation of $M_{\rm UV}$ only considers stellar
radiation and not nebular emission.  In particular, \lya~emission from
the \hii~region will be strong in these galaxies, perhaps having an
intrinsic equivalent width (EW) as high as 1500~\AA~\citep{Schaerer02,
  Johnson09, Dijkstra10}.  In a static ISM/IGM, \lya~photons would be
absorbed by the surrounding neutral medium.  However,
\citet{Dijkstra10} showed that a $\ga 50 \kms$ \hi~outflow allows for
10--20 per cent of the \lya~emission to be transmitted through a
neutral IGM, still resulting in a strong EW of
$\sim$100~\AA~\citep[see also][]{Verhamme06, Verhamme08}.  Inspecting
the most massive galaxy in our simulation at redshifts 12 and 7.3, we
find that the intrinsic $\log L_{{\rm Ly}\alpha} = (37.3, 40.3)
\unit{erg s}^{-1}$ in a sphere of radius $2r_{\rm vir}$, assuming that
all ionising radiation is absorbed by the nearby neutral IGM.  This is
comparatively higher than $\log L_{\rm 1500} = (36.0, 39.0) \unit{erg
  s}^{-1}$ from stellar emission alone.  At these respective
redshifts, the galaxy has a total mass of $2.7 \times 10^7 \Ms$ and
$6.8 \times 10^8 \Ms$ and a stellar mass of $1.5 \times 10^4 \Ms$ and
$3.5 \times 10^6 \Ms$.  Outflows are ubiquitous in high-redshift dwarf
galaxies, and we thus expect that a significant fraction of
\lya~emission would be transmitted through the IGM.  We intend to
follow-up the observable \lya~line profiles and strengths in a more
complete statistical sample of high-redshift galaxies presented in
\citet{Xu13}.

Even with the exquisite sensitivity of JWST, the typical reionising
sources will not be detectable.  Only galaxies with a SFR $> 0.1
\hsfr$ will be detected in 10$^6 \unit{s}$ exposures \citep{Pawlik13}.
However, surveys of gravitational lensed regions may uncover a
significant fraction of such galaxies, for instance, median
magnifications are between 2--15 in the CLASH fields
\citep{Bouwens12_CLASH, Bradley13}.  Searches with narrow-band filters
at a particular redshift around the \lya~$\lambda1215$,
\heii~$\lambda1640$, and H$\alpha$ $\lambda6536$ lines
\citep[e.g.][]{Johnson09, Paardekooper13} in gravitational lensed
fields may be the best strategy for detecting high-redshift dwarf
galaxies \citep[e.g.][]{Zackrisson12}.  If detected, these faint
galaxies would be highly unobscured because the majority of the
radiation would be propagating through ionised channels in the ISM and
surrounding IGM.  Would they constitute {\it a new galaxy class} that
only exist prior to cosmic reionisation, existing in large-scale
neutral regions of the universe embedded in their own \hii~region?
Because of their unique environment, they would have no direct
present-day analogue, and models that include radiative Population III
and II/I stellar feedback, creating local \hii~regions and outflows
are necessary to make solid predictions of the connection between
their intrinsic and observable properties.

%

\subsection{Comparison to previous work}

\subsubsection{Star formation rate and efficiency}

W12 found that the mean stellar metallicity and its distribution
function were good discriminates of a plausible star formation and
feedback implementation, avoiding the typical galaxy formation
simulation ``overcooling problem''.  Little is published on simulated
stellar metallicity data in the first galaxies \citep[however,
see][]{Ricotti05_Hist}, so we will compare the SFRs and efficiencies
to previous work.

\citet{Ricotti08} found a power law dependency between halo mass and
star formation efficiency with a slope of 1.5 and 2 in their weak and
strong feedback cases in their cosmological radiation hydrodynamics
simulations of early galaxy formation.  In their weak case, $\log
f_\star \simeq -3.0$ and $-1.5$ at halo masses of $10^7 \Ms$ and $10^8
\Ms$, respectively.  We see no such trend in $f_\star$ (see Figure
\ref{fig:stats}), where we find a decreasing trend at $M \la 10^{7.5}
\Ms$ with a large scatter because the outflows from the initial star
formation event temporarily suppresses any further star formation.  At
larger masses it slowly increases to 1--3 per cent at $M = 10^{8.5} \Ms$,
where we would expect it to plateau at higher halo masses.  It is wise
to disregard their power law fit at higher masses because it diverges,
and the star formation efficiency probably levels out at higher masses
\citep[see][]{Pawlik13}.

We find acceptable agreement with the low-mass haloes in the ``LW+RT''
simulation of \citet{Pawlik13} that considered most of the physics
included in our simulation with the exception of a model of Population
III star formation and feedback, metal cooling, and radiation
pressure.  However, they include the effects of stellar radiative
feedback in their work, which plays a significant role in regulating
star formation in these low-mass galaxies.  They find $\log f_\star
\simeq -3.0$, slightly lower than our results, which could come from
the lack of metal-line cooling, resulting in lower cooling rates and
thus star formation efficiencies.  On the other hand, there are
similarities between our results, in that, they have a similar scatter
and no dependence on halo mass.  Their values of $f_\star$ increase
rapidly to $\sim0.1$ when it can atomically cool, which is about a
factor of five higher than our results.  Lastly, they find similar gas
fractions $f_{\rm gas} \ga 0.1$ that are depressed in low-mass haloes.

\citet{Biffi13} analyzed a cosmological simulation that samples mostly
MC haloes (up to $7 \times 10^7 \Ms$ at $z=9$) and includes distinct
models of Pop III and Pop II/I star formation.  In their $z=9$ data,
they find a suppressed gas fraction of 2--5 per cent in haloes with $M
\sim 10^6 \Ms$, rising to 10 per cent in $\sim10^7 \Ms$ haloes.  We
find a similar suppression, but the haloes only recover to $f_{\rm
  gas} = 0.1$ in $10^8 \Ms$ haloes.  We suspect that their elevated
gas fractions result from the lack of radiative feedback.  Star
formation is suppressed below a halo mass scale of $5 \times 10^6
\Ms$, agreeing with our results.  In haloes with $M \ge 10^7 \Ms$,
they find stellar mass fractions ($M_\star/M_{\rm vir}$) between
$10^{-3}$ and $10^{-4}$.  The stellar mass fractions in our work in
these haloes are in the range $1-4 \times 10^{-4}$.  Their objects
with high star formation efficiencies most probably arises from the
higher gas fractions.  They also find a population of haloes with a
very small $f_\star = 10^{-6} - 3 \times 10^{-5}$, which is below our
stellar mass resolution of 1000 \Ms.


\subsubsection{UV Escape fraction}

In the past five years, theoretical work has favored high escape
fractions between $0.1-0.8$ from galaxies with total masses $M \la
10^9 \Ms$ \citep{Wise09, Razoumov10, Yajima11, Paardekooper13}.  We
also find high average \fesc~values of $\sim$0.5 but only in haloes
with $M \la 2 \times 10^7 \Ms$, which are then photo-suppressed.  The
UV escape fraction steadily decreases with increasing halo mass to a
luminosity-weighted average of 25 and 5 per cent for $10^8 \Ms$ and
$10^{8.5} \Ms$ haloes, respectively.

Why do we find lower escape fraction in low-mass atomic cooling haloes
than other recent studies?  To explain this, we can describe the
shortcomings of \citet{Wise09}, who extracted haloes from cosmological
simulations without radiative cooling, star formation, and feedback as
initial conditions.  These haloes had a gas fraction near the cosmic
mean, which then proceeded to monolithically collapse because
radiative cooling was immediately activated in their high-resolution
re-simulations.  This produced unrealistically strong starbursts in
their ten halo sample with a star formation efficiencies of 5--10 per
cent and $\fesc \sim 0.4$ for haloes with $M \ga 10^8 \Ms$.  For such
strong starbursts, we found that $\fesc \ga 0.5$ (see Figure
\ref{fig:fesc2} and Table \ref{tab:fesc2}).  Because the $M = 10^8 -
10^9 \Ms$ haloes in \citet{Razoumov10} and \citet{Yajima11} were the
smallest in their sample, they would have suffered the same
aberration, which is a general consequence of missing early phases of
galaxy formation, leading to the overcooling problem.  If the haloes
were allowed to form stars throughout their assembly, then the SFRs
and efficiencies would have been regulated over this period, leading
to more controlled star formation events and lower \fesc~values.
Analogous behavior for the escape fraction was seen in the metal
cooling only simulation of W12, where the most massive halo underwent
a catastrophic cooling event once it reached $\tvir \sim 10^4
\unit{K}$, resulting in a SFR an order of magnitude higher than the RP
simulation, which is presented here, that proceeded to reionise nearly
the entire simulation volume.  Recently using radiation hydrodynamics
simulations, \citet{Kimm14} have found that the time-averaged mean
escape fraction 10--15 per cent in high-$z$ dwarf galaxies hosted in
halos with $10^8 \le M/\Ms \le 10^{10.5}$, which combined with our
results may provide an accurate estimate of \fesc~in galaxies that
contribute substantially to reionisation.

Our results for $M \sim 10^7 \Ms$ haloes are similar to the results of
\citet{Paardekooper13} before their uniform UV background activates at
$z = 12$.  Then their average \fesc~values increase from $\sim$0.6 to
nearly unity at $z = 6$.  Similar evolution occurs in $10^8 \Ms$
haloes, increasing from $\sim$0.4 to 0.95.  We suspect that variations
in our analyses, where they consider all star clusters and whereas we
only consider young ($<20$ Myr) star clusters, have caused the
discrepancy between our two efforts.  As the haloes are
photo-evaporated by the UV background at later times in their
simulation, the UV stellar radiation from the older stars can escape
more easily through a lower neutral column density, causing higher
\fesc~values at later times.  They also see a similar but more abrupt
transition from MC to atomic cooling haloes as the primary source of
ionising photons as the former objects are photo-suppressed.

%

\subsubsection{Reionisation history}

The inclusion of galaxies that are liable to photo-suppression into a
reionisation model produces an extended reionisation history with the
ionised fraction increasing from $x = 0.2$ at $z = 14$ to $x = 0.5$ at
$z = 9$, and finally becoming completely reionised by $z \simeq 6.5$.
This history is a consequence of a halo mass dependent escape fraction
and SFRs, extending down to $10^{6.5} \Ms$ haloes that are gradually
photo-suppressed.  It is very similar to the one presented in
\citet{Haardt12}, but the evolution of \fesc~and the cosmic SFRs are
much different in nature.  Because galaxies can form in MC haloes,
non-negligible SFRs extend to $z \sim 15-30$, reducing the need for an
escape fraction of unity at $z > 12.5$ in their model.  Recently a few
groups \citep{Shull12, Kuhlen12, Robertson13} calculated a
reionisation history that was based on the latest HUDF data,
extrapolating the LF down to various limiting magnitudes, where they
considered various constant values of \fesc~and a redshift-dependent
escape fraction.  In a recent analysis of ancient stars in local dwarf
galaxies, \citet{Salvadori14} found a similar reionisation history as
the aforementioned works, but they concluded that galaxies in haloes
with $10^7 \le M/\Ms \le 10^8$ can reionise 50 per cent of a Milky Way
environment by $z \approx 8$.  Our reionization history is more
extended than these works, where the difference comes from
low-luminosity galaxies contributing more at higher redshifts.  Their
resulting $\tau_e$ is still consistent with the latest WMAP and Planck
data but on the low-end of the 1-$\sigma$ errors.

Our reionisation model uses a similar approach as \citet{Alvarez12}
with $f_\star$ and \fesc~being halo mass dependent, however, we use
values that are calibrated from our simulation with six mass ranges
instead of two.  Their idea of a low-mass galaxy population driving
early reionisation is plausible and is supported by our simulations.
Like the aforementioned works, their reionisation history is slightly
shorter than our results, mainly arising from their model only
considering haloes with $M > 10^8 \Ms$ as star-forming.  The general
trends, such as a peak in ionising emissivity and early-time and
late-time asymptotes in the luminosity-averaged escape fraction, are
found in our model, but clearly with different source parameters, the
details have changed.  Nevertheless, we find this type of reionisation
history to be most plausible, considering the findings in our
simulation presented in this paper.


\subsection{Caveats}

Our simulations include most of the relevant physical processes in
galaxy formation and its associated star formation and feedback, but
there are still some shortcomings in our work.  The simulation volume
is only 1 comoving Mpc$^3$, which misses the effects of large-scale
reionisation and rare peaks in the cosmological density field.  This,
however, does not diminish our findings of the properties of early
dwarf galaxies because stellar feedback is the dominant factor in
simulating realistic galaxies \citep[e.g.][]{Wise12_RP, Stinson13,
  Hopkins13_FIRE}.  The small volume does restrict our analysis to 32
dwarf galaxies and evaluating the galaxy properties independent of
time.  The approach of combining the data at all outputs misses any
evolution in the galaxy properties, but we have included the
time-dependent effects of photo-suppression in our semi-analytic
reionisation model.

We have not considered the effects of relative streaming velocities
($v_{\rm vel} \sim 30 \kms$ at $z \sim 1100$) between baryons and DM
that arise during recombination \citep{Tselia10}.  This phenomenon
only suppresses Pop III star formation in the smallest minihaloes with
$M \la 10^6 \Ms$ \citep{Tselia11, Greif11_Delay, Naoz12, OLeary12} and
should not significantly change our results because the circular
velocities of the galaxy host haloes are much larger than the
streaming velocity at $z < 20$.

On the topic of the escape fraction, there are a few processes that we
do not model.  First, we cannot capture the possibility of runaway
massive stars that may boost \fesc~as these stars travel into the
outskirts of the dwarf galaxy and into the IGM \citep{Conroy12}
because we model metal-enriched star formation at the stellar cluster
scale.  Second, we do not include the partial ionisation by X-ray
radiation from X-ray binaries \citep{McQuinn12, Power13, Fragos13} or
mini-quasars \citep[e.g][]{Ricotti04_Xray, Kuhlen05, Holley12,
  Grissom14}, which may contribute up to 10 per cent of the optical
depth to Thomson scattering.  Finally, we consider a Salpeter IMF
($dN/dM_\star \propto M_\star^\alpha$ with $\alpha = -2.35$) for
metal-enriched stars, whereas recent observations of ultra-faint dwarf
galaxies have suggested that the IMF slope is more shallow at $\alpha
= 1.2^{+0.4}_{-0.5}$ for Hercules and $\alpha = 1.3 \pm 0.8$ for Leo
IV \citep{Geha13}.  The shallower slope implies that their progenitors
hosted star formation that favors massive stars more than present-day
star formation.  \citet{Wise09} also found that a top-heavy IMF
increases the escape fraction by $\Delta \fesc = 0.27 \pm 0.17$ but
also suppresses star formation by $\sim$25 per cent in the MC haloes.
Considering a shallower IMF may change our results slightly in the
low-mass end that we investigated, but it is still uncertain under
which conditions the IMF becomes more top-heavy \citep[e.g., see
differing results in][]{Jappsen09, Smith09, Safranek14}.  On a similar
topic, we used a characteristic mass of $100 \Ms$ for our Pop III IMF,
which is somewhat higher than recent Pop III star formation
simulations that find characteristic masses around tens of
\Ms~\citep{Turk09, Stacy10_Binary, Greif11_P3Cluster, Greif12,
  Susa13}.  However, a recent paper by \citet{Hirano14} studied 100
instances of Pop III protostar evolution with axisymmetric radiation
hydrodynamics simulations that were initialised from a cosmological
simulation.  They found a wide range of possible stellar masses
$M_\star = 10-1000 \Ms$, and under this scenario, our choice of a Pop
III IMF would still be within the uncertainties gathered from such
simulations.

%
%
%
%
%

\section{Conclusions}
\label{sec:conclusions}

We present the characteristic properties and abundances of dwarf
galaxies at high-redshift and their absolute contribution to cosmic
reionisation.  To obtain our results, we use a cosmological radiation
hydrodynamics simulation that considers Pop II and III star formation
with a self-consistent transition with their radiative feedback
modeled with the radiation transport module \moray.  In a previous
paper (W12), we showed that the star formation history and stellar
population of the most massive dwarf galaxy in the simulation analysed
here agreed with the local dwarf galaxy metallicity-luminosity
relation.  We have further analysed this simulation that captures the
buildup of the first galaxies, starting with their Pop III
progenitors, focusing on their global properties, LF, UV escape
fraction, and role during reionisation, and the highlights of our work
are as follows:

\begin{enumerate}
\item Low-luminosity galaxies with stellar masses up to $3 \times 10^4
  \Ms$, SFRs of $10^{-3}~\hsfr$, and absolute UV magnitudes of --12
  are able to form in metal-line cooling (MC) haloes ($T_{\rm vir} \le
  10^4 \unit{K}$).  This usually occurs in one burst, which then
  suppresses any further star formation through stellar feedback, and
  star formation will recommence after sufficient gas has
  (re-)accreted into the potential well.
\item Gas fractions in the MC haloes have a large spread and
  are, on average, $\sim$5--7 per cent, where $\sim$0.3--3 per cent of
  this gas form stars.  In addition to internal suppression, they are
  subsequently photo-suppressed by an external radiation field.
\item The early dwarf galaxy LF flattens to $\sim$3 mag$^{-1}$
  Mpc$^{-3}$ at $M_{\rm UV} \ga -12$.  The galaxies in this plateau
  form in MC haloes usually with one star formation event that
  produces an initial magnitude around $M_{\rm UV} = -12$ and then
  increases as the stellar population ages.
\item The luminosity-weighted escape fraction decreases with halo mass
  with $\fesc \simeq 0.5$ in haloes with $M \le 2 \times 10^7 \Ms$,
  $\fesc \simeq 0.3$ in haloes with $2 \times 10^7 \le M/\Ms \le 2
  \times 10^8$, and $\fesc \simeq 0.05$ in larger haloes.  The escape
  fraction is highly time dependent and is correlated with the SFR
  with an average delay of $\sim$10 Myr.
\item The amount of ionising photons per unit mass escaping from the
  halo, i.e. $\fesc f_\star f_{\rm gas}$, shows little evolution with
  halo mass with a mean value of $10^{-3.6}$ over the mass range
  captured in our simulation.
\item Low-luminosity galaxies hosted in MC haloes propel the early
  epochs of reionisation, providing 75 per cent of the instantaneous
  emissivity at $z = 14$ when our reionisation model has an ionisation
  fraction of 20 per cent.  These faintest galaxies contribute nearly
  30 per cent of the ionising photon budget by $z = 6$.
\item Photo-suppression of low-luminosity galaxies leads to a
  photon-starved reionisation scenario by $z = 6$, agreeing with
  emissivities inferred from \lya~forest observations.  By utilizing
  calibrated galaxy properties in our reionisation model, we obtain an
  optical depth to Thomson scattering $\tau_e = 0.093$, agreeing with
  the latest WMAP and Planck results.  
\item The luminosity-weighted escape fraction and host halo mass
  smoothly decline and increase, respectively, with time, and we have
  given functional fits (Equations \ref{eqn:fesc} and \ref{eqn:mvir})
  to these trends for use in future studies.
\end{enumerate}

We have shown that the faintest galaxies contribute a significant
amount to the ionising photon budget during cosmic reionisation.
Their consideration in reionisation calculations is essential in order
to adhere to observational constrains, such as $\tau_e$, the duration
of reionisation, a mostly ionised IGM by $z \sim 6$, and the
ionisation background at $z = 4-6$.  We are currently following up
this study with a larger dataset with more massive galaxies
\citep{Xu13} to further constrain galaxy scaling relations and galaxy
observables during reionisation, which is timely with the future
launch of JWST and commissioning of 30-m class ground-based
telescopes.

\section*{Acknowledgments}

We thank an anonymous referee for an insightful review that helped
improve this manuscript.  JHW~has appreciated helpful conversations
with Marcelo Alvarez and Pascal Oesch and acknowledges support by NSF
grants AST-1211626 and AST-1333360.  MJT~acknowledges support by the
NSF CI TraCS fellowship award OCI-1048505.  Computational resources
were provided by NASA/NCCS award SMD-11-2258 and XSEDE allocation
AST-120046.  This work was also partially supported by NSF grant
AST-1109243. This research has made use of NASA's Astrophysics Data
System Bibliographic Services.  The majority of the analysis and plots
were done with \yt~\citep{yt_full_paper}.

\bibliography{ms}
\bsp
\label{lastpage}

\end{document}